\documentclass[11pt,a4paper]{article}
\usepackage{jheppub}
\usepackage[dvipsnames, svgnames, x11names]{xcolor}
\usepackage{lmodern}
\usepackage[utf8]{inputenc}
\usepackage{color}
\usepackage{float}
\usepackage{amsmath,bm}
\usepackage{amssymb}
\usepackage{graphicx}
\usepackage{setspace}
\usepackage{subfigure}
\usepackage{amsthm}
\usepackage{amsfonts}
\usepackage{braket}
\usepackage{multirow}
\usepackage{ulem}
\usepackage{slashed}
\usepackage{babel}
\usepackage{lipsum}
\usepackage{hyperref}
\usepackage{url}
\hypersetup{
	colorlinks=true,
	linkcolor=cyan,
	urlcolor=blue,
	citecolor=red,
}

\title{Entanglement islands of 2D charged scalar-hairy black holes on the brane}

\author[a]{Qingsong Li}
\author[a]{Yuxuan Liu}
\author[a]{Wenjie Zhu\note{Corresponding author.}}

\affiliation[a]{
Institute of Quantum Physics, School of Physics, Central South University, Changsha 410083, China
}

\emailAdd{liqingsong202606@163.com}
\emailAdd{liuyuxuan93@csu.edu.cn}
\emailAdd{zhuwenjie9978@163.com}

\abstract{
We investigate the entanglement dynamics and Page curve of a 2D brane Maxwell-Scalar (MS) black hole coupled to thermal baths in double holography. Computing the entanglement entropy via the island formula, we find that injecting electric charge and scalar source into the system produce opposite effects: the former increases the degrees of freedom (DOF) on the brane by enhancing the effective global tension, thereby raising the saturation entropy of the system; conversely, the latter reduces the DOF on the brane by lowering the effective local tension near the IR geometry, ultimately suppressing the entanglement. Despite the non-trivial backreaction from the matter fields, the pre-saturation entanglement dynamics remain robust. We identify an early-time quadratic growth corresponding to the pre-local-thermalization phase, which smoothly transitions into a late-time linear growth. 
Furthermore, the scale of the Page time is governed by a competition between thermal excitations and the brane DOF. Higher temperatures accelerate the onset of the Page time by exciting more Hawking modes, while abundant DOF delay it by increasing the saturated entropy. 
}

\begin{document}

\maketitle

\section{Introduction}
Understanding the fundamental physics of black holes through the lens of quantum mechanics remains a major challenge in modern theoretical physics. This endeavor inevitably confronts the black hole information paradox, which questions whether the evaporation process or the exchange of Hawking radiation preserves unitarity \cite{hawking1974black,hawking1975particle,hawking1976breakdown}. Semiclassical gravity calculations predict a monotonic growth of entanglement entropy between the black hole and its radiation, implying an ultimate loss of information. However, from the perspective of unitary quantum physics, the entanglement entropy of the radiation must eventually turn over and decay at late stages, tracing out the dynamical evolution depicted by the Page curve \cite{Page:1993wv,Page:2004xp,Page:2013dx}.

Significant progress in resolving this paradox was made within the framework of Gauge/Gravity duality \cite{penington2020entanglement,Almheiri:2019psf}. By extremizing the generalized gravitational entropy \cite{Lewkowycz:2013nqa,Engelhardt:2014gca}, the entanglement attributed to Hawking radiation can be elegantly captured by the quantum extremal surface (QES) and the island formula \cite{Almheiri:2019hni}:
\begin{equation}\label{eq:QESinRad}
S[\mathcal{R}]=\min_{\mathcal I} \left\{ \mathop{\text{ext}}\limits_{\mathcal I} \left[\frac{\mathbf{A}[\partial \mathcal{I}]}{4 G_{N}} + S[\mathcal{R} \cup \mathcal{I}]\right]\right\},
\end{equation}
where $\mathbf{A}$ represents the area, while $\mathcal{R}$ and $\mathcal{I}$ denote the radiation and the islands, respectively. The emergence of the island in the gravitational region during the late stages of evolution pinches off the unbounded growth of entanglement, yielding a result perfectly consistent with the Page curve \cite{Almheiri:2020cfm}.
Since its proposal, the island paradigm has been widely investigated across various lower-dimensional models and asymptotical boundaries \cite{Chen:2019uhq,Alishahiha:2020qza,Hashimoto:2020cas,Anegawa:2020ezn,Hartman:2020swn,Chen:2020jvn,Bhattacharya:2020uun,Deng:2020ent,Wang:2021woy,He:2021mst,Gautason:2020tmk,Krishnan:2020oun,Sybesma:2020fxg,Chou:2021boq,Hollowood:2021lsw,Suzuki:2022xwv,Suzuki:2022yru,Bhattacharya:2021nqj,Bhattacharya:2021dnd,Caceres:2021fuw,Bhattacharya:2021jrn,Caceres:2020jcn}. In the path integral formalism, the emergence of islands was demonstrated to correspond to the dominance of replica wormholes linking different replica manifolds \cite{Penington:2019kki,Almheiri:2019qdq,Rozali:2019day,Karlsson:2020uga}, which inspired discussions on disjoint universes and baby universes connected by wormholes \cite{Balasubramanian:2020xqf,Balasubramanian:2021wgd,Balasubramanian:2020coy,Miyata:2021ncm,Miyata:2021qsm,Marolf:2020rpm,Marolf:2020xie,Balasubramanian:2020jhl,Peng:2021vhs}. Furthermore, various quantum information measures, such as reflected entropy and mutual information, have been utilized to explore the mixed-state entanglement structures inside the gravity system and the radiation \cite{Renner:2021qbe,Li:2020ceg,KumarBasak:2020ams,Kawabata:2021hac,Kawabata:2021vyo,Vardhan:2021mdy,Akal:2021dqt,Ling:2021vxe,Chandrasekaran:2020qtn,Li:2021dmf,Czech:2023rbh,Liu:2023ggg,Liu:2026ruv}.

Beyond evaporating black holes, the information paradox also manifests in static geometries where an eternal black hole equilibrates with flat thermal baths maintained at a matched temperature, a process that can be interpreted as the evolution of the global thermofield-double state or coupled Sachdev–Ye–Kitaev clusters \cite{Almheiri:2019yqk,Maldacena:2001kr,Gu:2017njx,Hartman:2013qma,Chen:2020wiq,Liu:2022pan,Geng:2021iyq,Geng:2021mic,Afrasiar:2022ebi,Wang:2023vkq,Liu:2024cmv}. Despite the absence of net energy exchange, the entanglement between the black hole and the radiation is generated continuously. To evaluate the island formula non-perturbatively in such setups, one frequently employs the doubly holographic framework \cite{Almheiri:2019hni,Chen:2019uhq,Almheiri:2019psy}, which originates from the holographic dual of boundary conformal field theory \cite{Takayanagi:2011zk,Nozaki:2012qd}. In this paradigm, the black hole is realized on a finite-tension Planck brane subjected to Neumann boundary conditions, translating the QES into a standard Hubeny-Rangamani-Takayanagi (HRT) surface in the higher-dimensional bulk. Crucially, a prerequisite for realizing a dynamical Page curve in this scenario is that the brane must harbor a sufficient number of DOF~\cite{Ling:2020laa,Chen:2020uac,Geng:2020fxl,Geng:2020qvw}.

However, early doubly holographic models were primarily restricted to purely geometrical setups, modeling the brane with a constant tension or introducing localized Dvali-Gabadadze-Porrati (DGP) gravity. These models lacked dynamical localized matter DOF on the brane, precluding the existence of complex internal structures. Recently, this rigid structure has been significantly enriched by introducing novel dynamical fields onto the brane \cite{Hao:2025ocu,Fujiki:2025yyf,Shinmyo:2023eci,Kanda:2023jyi,Miao:2024olz,Cui:2023gtf,Guo:2023fly,Miao:2023mui}. 

Following these advancements, the physical aspect regarding black holes interacting with their environments becomes highly relevant. In reality, black holes act as open systems that inevitably interact with surrounding classical matter fields during the quantum emission of Hawking radiation. Consequently, it is expected that the presence of such matter fields will fundamentally affect the subsequent quantum entanglement dynamics and alter the Page curve.

In this paper, we model these phenomena by investigating the entanglement dynamics of an MS black hole localized on a brane. Specifically, we explore how the backreaction of scalar hair—acting as primary hair driven by environmental interactions—governs the entanglement exchange between the brane black hole and the thermal bath. By evaluating the holographic entanglement entropy, we aim to investigate the impact of the scalar on the dynamical phase structures, the saturation entropy, and the rate of entanglement propagation.

The paper is organized as follows: in Section \ref{sec:1}, we establish the doubly holographic framework and construct the action and equations of motion for the MS black hole on the brane. In Section \ref{sec:2}, we present the static geometric configurations of the brane and analytically investigate the Page curve in the bare Maxwell limit to establish a baseline. Subsequently, we take the full scalar backreaction into account by numerically solving the system, and explore the effect of scalar hair on the Page curve and its temperature-dependent phase structure. Finally, the conclusions and discussions are given in Section \ref{sec:3}.

\section{The setup in double holography}\label{sec:1}

\begin{figure} 
  \centering
  \subfigure[]{\label{fig:braneP}
  \includegraphics[height=0.25\linewidth]{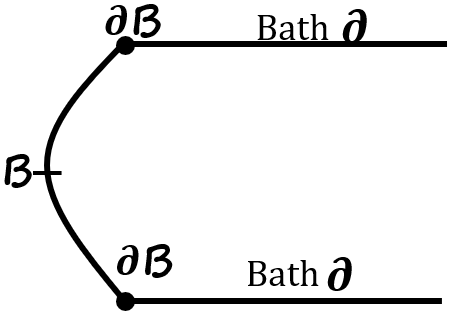}}
    \hspace{50pt}
  \subfigure[]{\label{fig:bulkP}
  \includegraphics[height=0.25\linewidth]{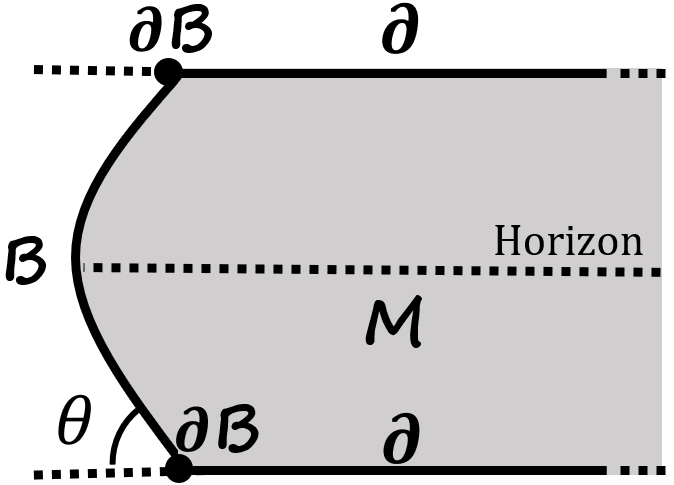}}
\caption{(a): Brane perspective. A $2$-dimensional eternal black hole $\mathcal{B}$ is in equilibrium with two flat baths $\bm{\partial}$. (b): Bulk gravity perspective. The Planck brane $\mathcal{B}$ is embedded in the ambient BTZ black holes.}
\end{figure}

Specify a composite system consisting of a (1+1)-dimensional black hole $\mathcal{B}$ and heat baths $\bm{\partial}$ that are coupled to its spacetime boundary $\partial\mathcal{B}$ -- Fig.~\ref{fig:braneP}. The temperature of the heat bath is the same as that of the black hole, such that the entire system remains in thermal equilibrium during the exchange of Hawking radiation. As shown in Fig.~\ref{fig:bulkP}, this setup admits a higher-dimensional holographic dual described by a BTZ black hole $\mathcal{M}$ inserted by a co-dimension-one Planck brane $\mathcal{B}$, with the brane meeting the conformal boundary $\bm{\partial}$ at an intersection $\partial\mathcal{B}$ \cite{Almheiri:2019hni,Chen:2020uac,Almheiri:2019yqk,Takayanagi:2011zk}. 

From this perspective, the action of the $(2+1)$-dimensional gravity theory is specified as
\begin{align}\label{eq:Action}
I=&\frac{1}{16\pi G_N} \Bigg[ \int_\mathcal{M} d^{3}x
\sqrt{-g}\left(R+2\right)+2\int_{\partial}d^{2}x\sqrt{-h_{\partial}}K_{\partial}\nonumber\\
&+2\int_{\mathcal{B}}d^{2}x\sqrt{-h_{\mathcal{B}}}K_{\mathcal{B}}\Bigg] - \int_\mathcal{B} d^2 x
\sqrt{-h}\mathcal{L}_M + \text{Junction terms on $\partial\mathcal{B}$}.
\end{align}
In this setup, $G_N$ denotes the Newton constant in the bulk. The quantities $K_{\mathcal{K}}$ and $h_{\mathcal{K}}$ represent the extrinsic curvature and the induced metric on the hypersurface $\mathcal{K}$, where $\mathcal{K}=\bm{\partial}$ denotes the conformal boundary and $\mathcal{K}=\mathcal{B}$ the Planck brane. Moreover, the Anti-de Sitter (AdS) radius is taken to be $L=1$ and $\mathcal{L}_M$ describes the matter fields localized on the brane. 

In this work, we investigate the information paradox for a class of MS black holes, which have been extensively studied in the context of holographic Kondo models \cite{Erdmenger:2013dpa,Erdmenger:2014xya,Erdmenger:2015spo,Erdmenger:2015xpq,Erdmenger:2016msd,Erdmenger:2020fqe}. The matter Lagrangian on the brane is taken to be
\begin{equation} 
    \mathcal{L}_M
    =\frac{1}{4}F_{ab}F^{ab}
    +D_a\Phi (D^a\Phi)^\dagger
    +m^2|\Phi|^2,
\end{equation}
where the index $a$ runs over directions along the brane, $F=dA$ denotes the field strength of the $U(1)$ gauge field $A$, and $\Phi$ is a complex scalar field minimally coupled to $A$. The gauge-covariant derivative is defined as $D_a=\nabla_a-i q A_a$.

\subsection{Ansatz of fields}
Since all matter fields are localized on the brane, the bulk equations of motion (EOMs) admit the vacuum solution given by
\begin{equation}
    R_{\mu\nu}+2 g_{\mu\nu}=0.
\end{equation}
The EOMs governing the brane embedding are
\begin{align}\label{eq:eomofinducedmetric}
    K_{ab}-K h_{ab}=\lambda T_{ab},
\end{align}
where $\lambda=8 \pi G_N$, and the stress-energy tensor of the brane-localized matter fields is
\begin{align}
    T_{ab}
    :=\frac{2}{\sqrt{-h}}\frac{\delta}{\delta h_{ab}}
    \left(\sqrt{-h}\mathcal{L}_M\right)
    =-h_{ab}\mathcal{L}_M + F_a^{\ c} F_{bc}
    +2 D_a\Phi (D_b \Phi)^\dagger .
\end{align}
The EOMs for the brane matter fields read as
\begin{align}
    D^a D_a \Phi &= m^2 \Phi, \label{eq:eomofscalar}\\
    \nabla_a F^{ab} &= q J^b , \label{eq:eomofgauge}
\end{align}
where the conserved Noether current associated with the scalar field is
\begin{equation}
    J^a = i\left[\Phi^\dagger D^a \Phi
    -\Phi (D^a\Phi)^\dagger\right].
\end{equation}

Restricting to static configurations, we parameterize the brane embedding as $X=X(z)$ in the Banados-Teitelboim-Zanelli (BTZ) background as
\begin{equation}
    ds^2=\frac{1}{z^2}
    \left(
    -f(z)\,dt^2+\frac{dz^2}{f(z)}+dx^2
    \right),
    \qquad
    f(z)=1-\frac{z^2}{z_h^2}.
\end{equation}
The induced metric on the brane then takes the form as
\begin{equation}\label{eq:inducedmetric}
    ds^2=\frac{1}{z^2}
    \left[
    -f(z)\,dt^2+
    \left(\frac{1}{f(z)}+X'(z)^2\right)dz^2
    \right].
\end{equation}
For the matter sector, we also impose a static ansatz as
\begin{equation}\label{eq:matterfields}
    \Phi=\phi(z),
    \qquad
    A=A_a dx^a = A_t(z)\,dt .
\end{equation}

After the substitution of (\ref{eq:inducedmetric}) and (\ref{eq:matterfields}) into  (\ref{eq:eomofinducedmetric}), (\ref{eq:eomofscalar}) and (\ref{eq:eomofgauge}), we finally obtain three EOMs of $X(z), \phi(z)$ and $A_t(z)$, which are so lengthy that we will not present them here.

\subsection{Asymptotic behavior}
The composite system admits an analytic solution when the scalar field is turned off, $\phi(z)=0$. In this scenario, the brane EOM (\ref{eq:eomofinducedmetric}) reduces to the ordinary differential equation given by 
\begin{equation}
    z_h^2 \,X''(z)+z\, X'(z)^3 =0.
\end{equation}
Near the conformal boundary, the geometric interpretation of the brane embedding imposes the asymptotic expansion as
\begin{equation}\label{eq:asyx1}
    X(z) \sim  - \cot \theta \,z + \mathcal{O}(z^3) ,
\end{equation}
where the anchoring position of the brane on the boundary may be set to $(x,z)=(0,0)$ without loss of generality, while $\theta$ denotes the exterior angle at which the brane intersects the boundary.

Solving the above equation yields the analytic brane profile
\begin{equation}\label{eq:analyticx}
    X(z)=\frac{1}{2} z_h \log \left(\frac{\sqrt{z^2+z_h^2 \tan ^2\theta}-z}{\sqrt{z^2+z_h^2 \tan ^2\theta}+z}\right).
\end{equation}
With the scalar field turned off, the gauge-field EOM \eqref{eq:eomofgauge} admits a conserved first integral,
\begin{equation}
    z^2 A_t'(z)=\sqrt{1+f(z)X'(z)^2} \bm C,
\end{equation}
where $\bm C$ is an integration constant. Integrating over $z$, we further have
\begin{equation}\label{eq:analyticA}
    A_t(z)=A_t(z_h)+\frac{\bm{C} \cot \theta \csc \theta \left(z \sec \theta-\sqrt{z^2+z_h^2 \tan ^2\theta}\right)}{z z_h}.
\end{equation}
The integration constant $\bm{C}$ is fixed by imposing regularity at the event horizon as
\begin{equation}
    \lambda  z_h^4 A_t'(z_h)^2+2 X'(z_h)=0 \quad \text{and} \quad A_t(z_h)=0.
\end{equation}
Substituting \eqref{eq:analyticx} and \eqref{eq:analyticA} into these conditions yields $$\bm{C}=\sqrt{\frac{2 \cos \theta}{\lambda}}.$$ It indicates a leading divergence of the gauge fields near the boundary as
\begin{equation}\label{eq:asya1}
    A_t \sim -\frac{\sqrt{2 \cos \theta}\csc \theta}{\sqrt{\lambda}z} + \cdots=: -\frac{Q}{z}+ \cdots.
\end{equation}
We note that for space dimension $d\neq 2$, the asymptotic behavior of the gauge field takes the standard form as
\begin{equation}
    A_t(z)\sim -\rho\, z^{d-2}+\mu,
\end{equation}
where $\rho$ and $\mu$ are identified as the charge density and chemical potential.

When turning on the scalar $\Phi$, we assume the near-boundary behavior $\phi(z)\sim z^\Delta$ and substitute this ansatz into \eqref{eq:eomofscalar} to determine the scaling dimensions $\Delta$. We further require that the scalar backreaction does not modify the leading asymptotic behaviors of $X(z)$ and $A(z)$ given in \eqref{eq:asyx1} and \eqref{eq:asya1}. Under these conditions, the scalar admits the expansion without a notorious logarithmic term as 
\begin{equation}
    \phi(z) = \phi_s z^{\Delta_-} + \phi_v z^{\Delta_+} +\cdots
\end{equation}
with $\Delta_-=1/4, \Delta_+=3/4$, and the parameters being fixed as
\begin{equation}\label{eq:parameter}
    Q=\sqrt{\frac{2\cot \theta \csc \theta}{\lambda}},\quad \text{and} \quad m^2= \frac{1}{16} \left(\sin ^2\theta+2 \cos 2 \theta +\frac{32 q^2 \cot \theta \csc \theta}{\lambda}-2\right).
\end{equation}

In conclusion, the near-boundary $(z\to 0)$ asymptotic expansions take the form as
\begin{align}\label{eq:bcs0}
    X(z) &= - \cot \theta z + \cdots, \nonumber\\
 \phi(z) &= \phi_s z^{1/4} + \phi_v z^{3/4} +\cdots, \\
 A(z)    &= -\frac{Q}{z} + \cdots,\nonumber
\end{align}
while regularity at the horizon $z=z_h$ requires
\begin{align}\label{eq:bcs1}
       X'(z_h) &= -\frac{1}{2}\left(3 z_h^4 A'(z_h)^2+2 m^2 \phi(z_h)^2\right),\nonumber\\
 \phi'(z_h) &= -\frac{m^2 \phi(z_h)}{2z_h}, \\
 A(z_h)    &= 0.\nonumber
\end{align}

We have thus reduced the brane black hole system with scalar hair to a coupled set of ordinary differential equations supplemented by the boundary conditions \eqref{eq:bcs0} and \eqref{eq:bcs1}. The resulting static configurations can be obtained numerically, as we discuss in Sec.~\ref{sec:num}.

\subsection{The entanglement entropy}
From the brane perspective, the brane black hole $\mathcal{B}$ is coupled to an external bath $\partial$ at its asymptotic infinity. Since Hawking radiation $\mathcal{R}$ emitted by the black hole is ultimately absorbed by the bath, the entanglement entropy of the radiation $\mathcal{R}$ can be computed via a QES anchored at the intersection $\partial\mathcal{B}$ between the black hole and the bath. As argued in \cite{Almheiri:2019hni}, to fully capture the DOF associated with the black hole, the brane subsystem must be extended to include a finite interval $[0, x_0]$ in the adjacent bath. Therefore, we partition the total system by identifying the region $x \leq x_0$ as the effective brane subsystem, and the remaining region $x > x_0$ as the radiation subsystem $\mathcal{R}$.

In light of the Island rule, the entanglement entropy of the radiation $\mathcal{R} $ is determined by a HRT surface in the bulk $\mathcal{M}$, given as \cite{Almheiri:2019psy,Chen:2020uac,Chen:2020hmv,Hernandez:2020nem}
\begin{align}\label{eq:SR}
S[\mathcal R]=\frac{1}{4G_N} \min_{\mathcal I} \left\{\mathop{\text{ext}}\limits_{\mathcal I} \left[\textbf{A}(\gamma_{\mathcal I\cup\mathcal R})\right]\right\},
\end{align}
where the entanglement island $\mathcal{I}$ lies on the brane $\mathcal{B}$, and $\gamma_{\mathcal I\cup\mathcal R} $ represents a codimension-two surface that shares its boundary with $\mathcal I\cup\mathcal R $. Moreover, the candidates HRT surfaces $\hat\gamma_{\mathcal I\cup\mathcal R} $ is obtained via extremizing over all possible islands, while the genuine HRT surfaces $\tilde\gamma_{\mathcal I\cup\mathcal R}$ can be obtained by choosing the configuration $\hat\gamma_{\mathcal I\cup\mathcal R} $ with minimal area $\tilde{\textbf{A}}$.

\begin{figure}
  \centering
  \subfigure[]{\label{fig:gammac}
  \includegraphics[height=0.33\linewidth]{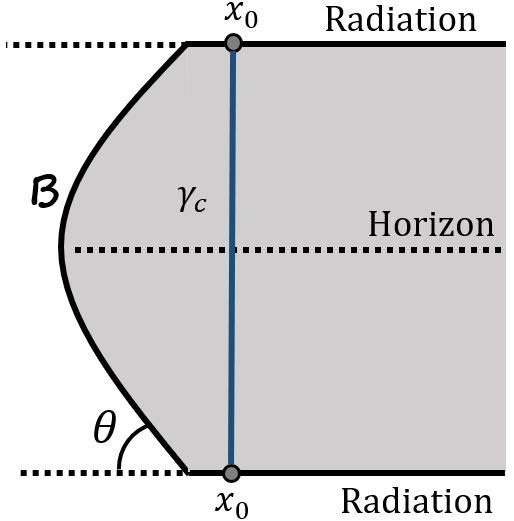}}
    \hspace{50pt}
  \subfigure[]{\label{fig:gammad}
  \includegraphics[height=0.33\linewidth]{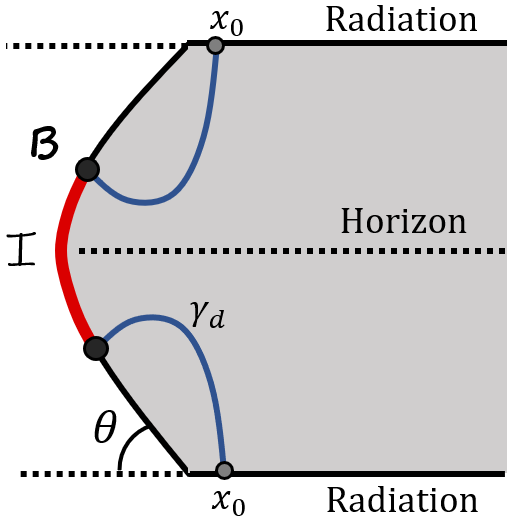}}
\caption{The configurations of HRT surfaces in two possible profiles: (a): the connected surface $\gamma_c$;
(b): the disconnected surface $\gamma_d$, with the HRT surfaces being plotted in blue.}\label{fig:gamma}
\end{figure}

In the present model, there are two candidate HRT surfaces -- Fig.~\ref{fig:gamma}. 

\paragraph*{Connected surface\\}
The first candidate is the \textbf{connected surface} $ \hat\gamma_{c} $ -- Fig.\ref{fig:gammac}, which is anchored to both the left and right baths, and the island $\mathcal{I}$ is absent. This surface grows with time and corresponds to the increasing entropy \eqref{eq:SR} due to the accumulation of the Hawking radiation. 

The extremal surface $\gamma_{c}=\hat{\gamma}_c$ can be parameterized as $$\Xi^\mu(\sigma)=\{t(\sigma),z(\sigma),x=x_0\},$$ with boundary conditions $$\{t=t_L=t_R,x=x_0,z=0\}$$ on the both Left and Right intersections.  The corresponding area formula can be expressed as
\begin{equation}\label{eq:confunctional}
    \textbf{A}(\hat\gamma_{c})=\int d\sigma \mathcal{L}_c=\int  \frac{d\sigma}{z}\sqrt{\frac{z'^2-f(z)^2 t'^2}{ f(z)}}.
\end{equation}
Choosing $\sigma$ as an affine parameter, we can impose the normalization condition as
\begin{equation}\label{eq:normalization}
    z'^2-f(z)^2 t'^2 = z^2 f(z).
\end{equation}
Since the metric is static, $t$ is a Killing direction, and the associated conserved quantity is
\begin{equation}\label{eq:conserved}
    E=\frac{\partial \mathcal{L}_c}{\partial t'} = -\frac{f(z) t'}{z^2}.
\end{equation}
Substituting \eqref{eq:conserved} into \eqref{eq:normalization}, we obtain 
\begin{equation}\label{eq:normalization2}
    z'^2= E^2 z^4 + f(z) z^2.
\end{equation}

Define the turning point $z_t$ of $\hat\gamma_c$ by
\begin{equation}
    z'|_{z=z_t}=0,
\end{equation}
which implies
\begin{equation}\label{eq:Eexpression}
    E^2=-\frac{f(z_t)}{z_t^{2}}.
\end{equation}
For the \textbf{connected surface} $\hat\gamma_c$, this turning point lies behind the horizon as $z_t>z_h$, so $f(z_t)<0$, and hence $E^2>0$. 

Substituting \eqref{eq:Eexpression} back into \eqref{eq:normalization2}, we find
\begin{equation}\label{eq:normalization3}
    z'^2= -\frac{f(z_t)}{z_t^{2}} z^4 + f(z) z^2.
\end{equation}

Next, combining \eqref{eq:conserved} and \eqref{eq:normalization3}, we obtain
\begin{equation}
\frac{dt}{dz}
=
\frac{t'}{z'}
=
\frac{E z^2 / f(z)}
{z\sqrt{
f(z)-\frac{z^2}{z_t^2}f(z_t)
}}.
\end{equation}
The boundary time separation is therefore
\begin{equation}
t
=
\int_0^{z_t}
dz\,
\frac{E z}
{f(z)\sqrt{
f(z)-\frac{z^2}{z_t^2}f(z_t)
}}.
\label{eq:timeIntegral}
\end{equation}
For $f(z)=1-\frac{z^2}{z_h^2}$,
the integral \eqref{eq:timeIntegral} can be evaluated analytically, yielding (App.~\ref{app_bdyt})
\begin{equation}
\cosh\left(\frac{t}{z_h}\right)=\frac{z_t}{z_h}.
\label{eq:time}
\end{equation}

The area formula \eqref{eq:confunctional} thus reduces to
\begin{align}\label{eq:confunctional-h}
    \textbf{A}(\hat\gamma_{c})=\int_\epsilon^{z_t}  \frac{dz}{z'}=\log\left(\frac{2 z_h}{\epsilon}\cosh\left(\frac{t}{z_h}\right)\right),
\end{align}
which is consistent with \cite{Maldacena:2013xja}. By analyzing this formula, we can identify two asymptotic behaviors: at early times, the entropy exhibits quadratic growth,
\begin{equation}\label{eq:earlygrowth}
    \textbf{A}(\hat{\gamma}_c)\bigg|_{t \ll z_h} \simeq \log\frac{1}{\pi T_h\epsilon} + 2\pi^2T_h^2t^2.
\end{equation}
In this regime, the brane and bath have just begun to exchange the Hawking modes, and the correlation is built up locally.
while at late times, the entropy grows linearly,
\begin{equation}\label{eq:lategrowth}
    \textbf{A}(\hat{\gamma}_c)\bigg|_{t \gg z_h} \simeq \log\frac{1}{\pi T_h\epsilon} + 2\pi T_h t.
\end{equation}
This linear behavior reflects the ballistic spreading of entanglement governed by the Hawking temperature~\cite{Liu:2013iza,Liu:2013qca,Leichenauer:2015xra}, where entanglement propagates at a constant velocity $v_E=2\pi T_h$.

\paragraph*{Disconnected surface\\}
The second candidate is the \textbf{disconnected surface} $\hat\gamma_{d}$ -- Fig.~\ref{fig:gammad}, which is also anchored on the intersection at $(x,z)=(x_0,0)$, but ends on the brane $\mathcal{B}$, giving rise to nontrivial islands $\mathcal I$. 
The presence of island $ \mathcal{I} $ ensures that the entropy \eqref{eq:SR} remains finite after the Page time \cite{Almheiri:2019yqk}.

In parallel, a general $\gamma_d$ can be parameterized by $x=x(z)$ and $t=0$. Therefore, the corresponding area functional is
\begin{equation}\label{eq:disfunctional}
    \textbf{A}(\gamma_{d})=\int dz \mathcal{L}_d=\int\frac{dz}{z}\sqrt{\frac{f(z) x'(z)^2+1}{f(z)}}.
\end{equation}
Due to the cyclic coordinate $x$ in \eqref{eq:disfunctional}, the corresponding conserved quantity can be obtained as
\begin{equation}\label{eq:conserved2}
  P=\frac{x'(z)}{z \sqrt{\frac{1}{f(z)}+x'(z)^2}}.
\end{equation}
At the turning point $z_*$ of $\gamma_d$, one has
\begin{equation}\label{eq:conserved3}
    x'(z_*)\to \infty, \quad \text{and} \quad P \to \frac{1}{z_*}.
\end{equation}
Combining the relations \eqref{eq:conserved2} and \eqref{eq:conserved3}, the profile of $x(z)$ is
\begin{equation}\label{eq:braneprofile}
    x(z)=x_0+k\; z_h \tanh^{-1}\left(\sqrt{\frac{z^2-z_*^2}{z^2-z_h^2}}\right)-z_h \tanh ^{-1}\left(\frac{z_*}{z_h}\right),
\end{equation}
where $k=1$ corresponds to the right branch with $x(z)\geq x(z_*)$, and $k=-1$ to the left branch with $x(z)\leq x(z_*)$.

Let $p_\mathcal{I}=(x_p,z_p)$ denote one boundary of the possible island, defined as the intersection of the surface $x(z)$ and the brane $X(z)$. The area of the surface $\gamma_d$ is therefore
\begin{align}\label{eq:disarea}
    \textbf{A}(\gamma_{d})
    &=  \left(\int_\epsilon^{z_*}+\int_{z_p}^{z_*}\right)\frac{dz}{z}\sqrt{\frac{f(z) x'(z)^2+1}{f(z)}},\nonumber\\
    &=\tanh^{-1}\left( \frac{z_h}{z_*} \sqrt{\frac{z_*^2-\epsilon^2}{z_h^2-\epsilon^2}} \right) + \tanh^{-1}\left( \frac{z_h}{z_*} \sqrt{\frac{z_*^2-z_p^2}{z_h^2-z_p^2}} \right),
\end{align}
which is consistent with the result in \cite{Sully:2020pza}.

At any possible island, one has the intersection condition as
\begin{equation}\label{eq:bcsI}
    x(z_p)=X(z_p)=x_p,
\end{equation}
In principle, the condition can be obtained by running over all possible island $p_\mathcal{I}$, and taking the configuration $\hat{x}(z)$ with locally minimal area $\textbf{A}(\hat{\gamma}_d)$. 
However, a more elegant approach follows from the variational principle, inspired by \cite{Ling:2021vxe,Liu:2023ggg}. We consider variations of the surface $$x(z)\to x(z)+\delta x(z)$$ and the island $$(x_p,z_p) \to (x_p + \delta x_p, z_p + \delta z_p),$$ then the variation of the area functional takes the form as
\begin{equation}
\delta \textbf{A}(\gamma_{d})
= \int dz \left( \frac{\partial \mathcal{L}_d}{\partial x} 
- \frac{d}{dz} \frac{\partial \mathcal{L}_d}{\partial x'} \right)\delta x
+ \left. \frac{\partial \mathcal{L}_d}{\partial x'} \delta x \right|_{z=z_p} + \left.\mathcal{L}_d \right|_{z=z_p} \delta z_p.
\end{equation}
The first bulk term vanishes upon imposing the Euler--Lagrange equation, leaving only the last two boundary contributions as
\begin{align}
    \delta \textbf{A}(\gamma_{d})_\text{bdy}&=\frac{\partial \mathcal{L}_d}{\partial x'} \delta x(z_p) + \left.\mathcal{L}_d \right|_{z=z_p} \delta z_p \nonumber \\
    &= \left[\frac{\partial \mathcal{L}_d}{\partial x'} (X'(z_p)-x'(z_p))+\mathcal{L}_d\right]\delta z_p.
\end{align}
The second equality arising from the fact that when the island $(x_p,z_p)$ is constrained to move along the brane trajectory $x_p=X(z_p)$, the profile $x(z)$ of $\gamma_d$ varies simultaneously. Therefore, the variation satisfies (\ref{app_vi})
\begin{equation}
\delta x(z_p) =    \left[X'(z_p)-x'(z_p)\right]\, \delta z_p,
\end{equation}
where $(x_p,z_p)$ denotes the boundary of the island, and $x(z_p)$ is the endpoint of $\gamma_d$ on the brane.
Requiring $\delta \textbf{A}(\gamma_d)=0$, we then find the Neumann boundary condition on the brane can be obtained as
\begin{equation}\label{eq:bcs2}
\frac{1}{f(z_p)} + x'(z_p) X'(z_p) = 0.
\end{equation}

It is worthy to note that this condition admits a simple geometric interpretation. 
Let $v_{d}^a$ and $v_{\mathcal{B}}^a$ denote the tangent vectors to the surface $\gamma_d$ and the brane respectively as
\begin{equation}
v_{d}^a = (1, x'), \qquad v_{\mathcal{B}}^a = (1, X').
\end{equation}
Then the above condition is equivalent to a transversality condition as
\begin{equation}
h_{ab} \, v_{d}^a \, v_{\mathcal{B}}^b = 0,
\end{equation}
namely that the two surfaces intersect orthogonally.

Combining \eqref{eq:bcsI} and \eqref{eq:bcs2}, we find that there is actually only one independent DOF among $z_p$, $z_*$ and the inclination angle $\theta$. Then, the extremal surface $\hat{\gamma}_d$ can be fully determined.
Finally, the entanglement entropy \eqref{eq:SR} is obtained by selecting the candidate with globally minimal area as
\begin{equation}
    S[\mathcal R]=\frac{1}{4G_N} \min \left\{\textbf{A}(\hat \gamma_c),\textbf{A}(\hat \gamma_d)\right\}.
\end{equation}

\section{Static Solutions on the Brane}\label{sec:2}
In this section, we will consider that the gravity-plus-bath system stays in thermal equilibrium, and analyze the influence of the classical matter fields on the brane black hole on the process of Hawking radiation.
\subsection{Configuration of the brane}\label{sec:num}

\begin{figure}
  \centering
  \subfigure[]{\label{fig:sXQ}
  \includegraphics[height=0.45\linewidth]{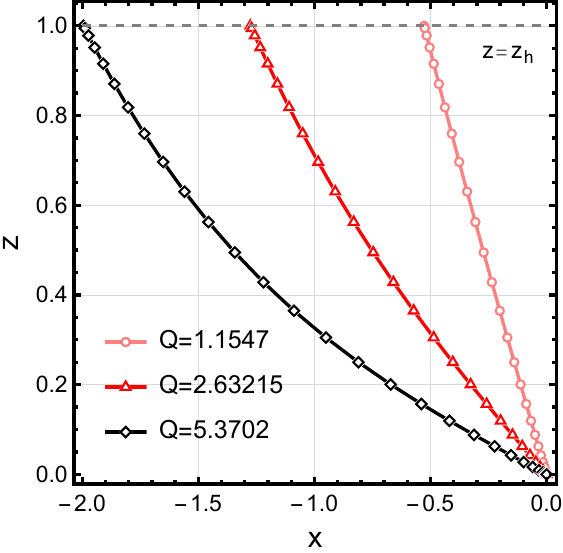}}
    \hspace{0pt}
  \subfigure[]{\label{fig:sXp}
  \includegraphics[height=0.45\linewidth]{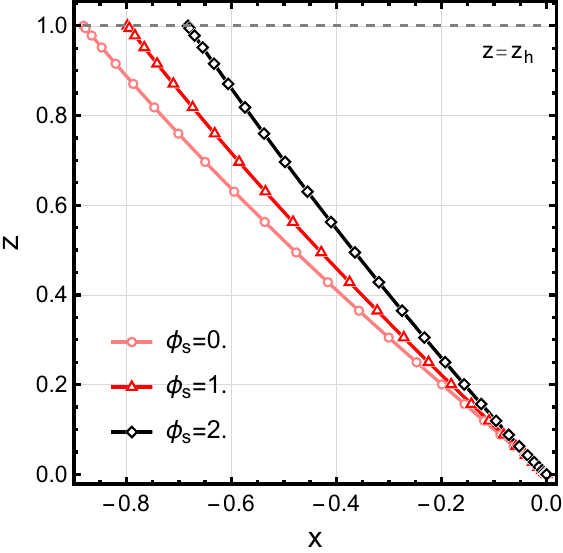}}
\caption{The numerical profiles of the brane embedding $X(z)$ at a fixed Hawking temperature $T_h=0.159$, together with (a): fixed scalar source $\phi_s=0.5$, but varying  charge $Q$, or (b): fixed electric charge $Q=1.682$, but varying source $\phi_s$.}
\end{figure}

Since there is no intrinsic gravitational contribution on the brane (such as the Jackiw-Teitelboim  gravity in $2$ dimensions, or the DGP term in higher dimensions), the brane matter fields on Hawking radiation will just change the configuration of the brane \cite{Chen:2020uac,Geng:2020qvw,Ling:2020laa}.

When turning off the scalar fields, the configuration of the brane has already been obtained in \eqref{eq:analyticx}. In contrast, when turning on the scalar, the system can only be solved numerically with the boundary conditions \eqref{eq:bcs0} and \eqref{eq:bcs1}. Specifically, we adopt the Chebyshev-Lobatto collocations and the Newton-Raphson iteration to solve the entire system. 

The input parameters of the system are $\{\lambda, \theta, q, \phi_s, z_h\}$. In the following, we instead work with the corresponding physical parameters $\{G_N,Q, m^2, \phi_s,T_h\}$. Throughout this paper, we fix $\lambda=1$ and $q=1/10$, such that the remaining independent parameters are $\{T_h, Q, \phi_s\}$.

As the electric charge $Q$ increases, the brane approaches the conformal boundary -- Fig.~\ref{fig:sXQ}. In contrast, increasing the scalar source $\phi_s$ drives the brane in the opposite direction -- Fig.~\ref{fig:sXp}. As discussed in \cite{Ling:2020laa,Liu:2022pan,Liu:2023ggg}, a smaller inclination angle typically corresponds to a larger brane tension, implying more DOF localized on the brane and hence a larger entanglement entropy. In the 3D model, despite the absence of an explicit tension term on the brane, the term arsing from the gauge field serves as an effective tension, given that
\begin{equation}
\frac{1}{4}F_{ab}F^{ab} \sim \cos \theta \quad \text{as} \quad z\to0,
\end{equation}
which exhibits the exact same asymptotic behavior as a global tension term.

Intriguingly, increasing the scalar source $\phi_s$ leads to a reduction in the effective DOF on the brane by bending the brane geometry deeper into the bulk, rather than by explicitly lowering the global effective tension. This behavior can be understood from the effective mass of the scalar field, which can be extracted from the linear term of $\Phi$ in \eqref{eq:eomofscalar} as
\begin{equation}\label{eq:effmass}
    M^2=m^2-\frac{q^2z^2A(z)^2}{f(z)} \sim -\frac{3}{16}\sin^2\theta \quad \text{as } z\to 0.
\end{equation}
The negative effective mass induces a negative contribution to the effective potential, $V(\Phi)\sim M^2 \Phi^2<0$, and only lowers the local tension of the brane in the deep IR as $z\to z_h$. Consequently, the IR geometry undergoes a deeper bending, which dynamically suppresses the effective DOF localized on the brane.

\subsection{Page curve}
In the current framework, the brane gravity $\mathcal{B}$ is coupled to the bath $\bm{\partial}$ eternally. Hawking radiation emitted by the black hole on the brane is continuously exchanged with the bath DOF. Therefore, the entanglement entropy between the bath $\bm{\partial}$ and the black hole $\mathcal{B}$ initially grows according to the connected surface \eqref{eq:confunctional-h}. Finally, when all DOF become entangled, the entanglement entropy saturates at \eqref{eq:disfunctional}, producing to a Page curve.

\subsubsection{2d Maxwell black holes on the brane}
\begin{figure}
  \centering
    \subfigure[]{\label{fig:saATheta}
  \includegraphics[height=0.45\linewidth]{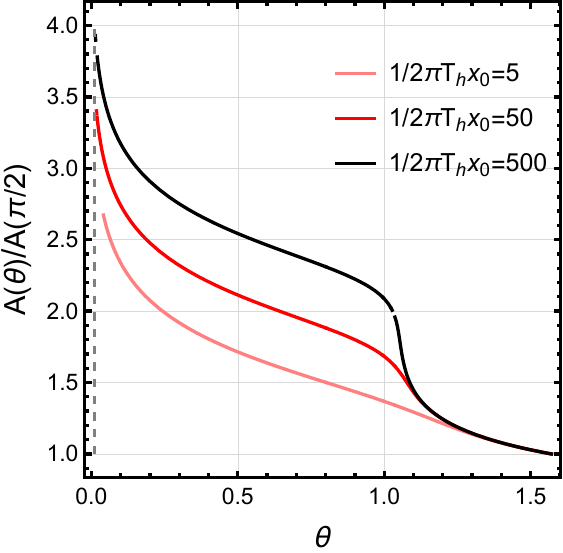}}
    \hspace{0pt}
\subfigure[]{\label{fig:saZsTheta}
  \includegraphics[height=0.45\linewidth]{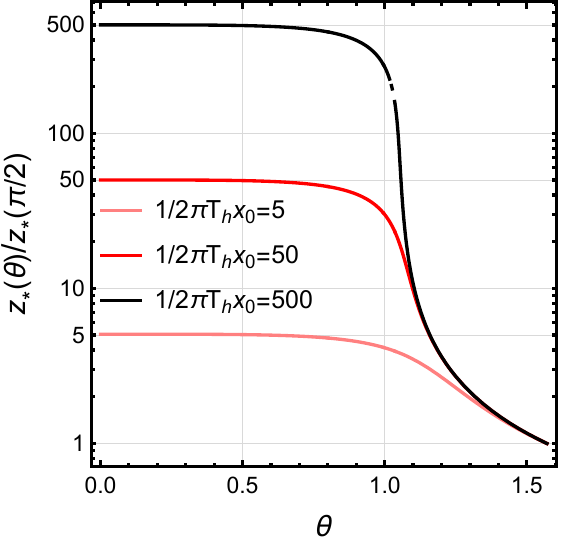}}
\caption{The analytical solutions with $\phi_s=0$ of (a): the area ratio of the disconnected extremal surface $\textbf{A}(\hat{\gamma_d})$,  and (b): the turning point ratio $z_*(\theta)$, both normalized by their value at $\theta=\pi/2$, for varying the inverse of Hawking temperatures $1/2\pi T_hx_0$.}
\end{figure}

When turning off the scalar field $\Phi$, the brane profile is described by \eqref{eq:braneprofile}, and thus, the transversality condition \eqref{eq:bcs2} implies
\begin{equation}\label{eq:island}
    z_p^2=\frac{z_*^2 \sin ^2\theta }{1-\frac{z_*^2 \cos ^2\theta }{z_h^2}}.
\end{equation}
Substituting this relation into the second term on the RHS of \eqref{eq:disarea}, one obtains
\begin{align}
        \textbf{A}(\hat{\gamma}_d)&=\tanh^{-1}\left( \frac{z_h}{z_*} \sqrt{\frac{z_*^2-\epsilon^2}{z_h^2-\epsilon^2}} \right) + \tanh^{-1}\left(\cos\theta\right)\nonumber\\
        &=\ln \frac{2z_* z_h/\sqrt{z_h^2-z_*^2}}{\epsilon}+  \ln \left(\cot \frac{\theta}{2}\right)+\mathcal{O}(\epsilon^2),
\end{align}
where $$z_*=  \frac{2 z_h\alpha(\theta)}{1+\alpha(\theta)^2}$$ can be fully determined by $\theta$ according to the expression $\alpha(\theta)$ defined in Sec.~\ref{app_para}. 
In the low-temperature regime, both the area ratio and the turning-point ratio exhibit a remarkably steep descent in the intermediate $\theta$ region -- Fig.~\ref{fig:saATheta} and Fig.~\ref{fig:saZsTheta}. This ``stretching'' behavior signifies a rapid crossover between two distinct geometric configurations of the extremal surface:
\begin{itemize}
    \item \textbf{The UV-Dominant Regime ($\theta \to \pi/2$)} In this regime, the geometry is nearly independent of the black hole size. All curves eventually converge to unity as $\theta \to \pi/2$, where the brane becomes a vertical probe and the extremal surface $\hat{\gamma}_d$ reduces to the standard semi-circle in the pure AdS limit.
    \item \textbf{The Horizon-Hugging Regime ($\theta \lesssim 1.0$)} In this regime, the extremal surface is strongly influenced by the IR geometry. The transversality condition forces the turning point $z_*$ to lie close to the horizon. Therefore, the area is also dominated by the near-horizon contributions.
\end{itemize}

In the neutral limit as $\theta \to \pi/2$ (or equivalently $Q \to 0$), one finds 
\begin{equation}
    \alpha \to \tanh \frac{x_0}{2 z_h},  \quad z_* \to z_h \tanh \frac{x_0}{z_h}.
\end{equation}
Thus, the area \eqref{eq:disarea} then reduces to
\begin{equation}\label{eq:disAweak}
    \textbf{A}(\hat{\gamma}_d)\simeq\ln \frac{\sinh (2\pi\, T_h\,x_0)}{\pi \,T_h\,\epsilon},
\end{equation}
which reproduces the classical entanglement entropy formula for the half-line in 2D CFT at finite temperature.

In the charged limit as $\theta \ll 1$ (or equivalently $Q \to \infty$) \cite{Chen:2020uac,Liu:2023ggg}, one instead obtains 
\begin{equation}
    \alpha \to 1 - \frac{1}{2}e^{-x_0/z_h} \theta ^2,  \quad z_* \to z_h \left(1-\frac{1}{8}e^{-2x_0/z_h}\theta^4\right).
\end{equation}
Correspondingly the area \eqref{eq:disarea} becomes
\begin{equation}\label{eq:disAsemi}
    \textbf{A}(\hat{\gamma}_d)\simeq 2\pi\, T_h\, x_0+\ln \frac{1}{\sqrt{2}\pi T_h\epsilon}+3 \ln Q.
\end{equation}
Two remarks are in order. First, the term in \eqref{eq:disAsemi} reflects a volume-law entanglement of the interval $x\in (0, x_0]$, as the charged limit drives the turning point $z_*$ toward the horizon $z_h$. A similar contribution $x_0/z_h \text{ with } 1/z_h=2\pi T_h$, also arises in the neutral limit \eqref{eq:disAweak}, as the endpoint $x_0$ extends deep into the bath $x_0\gg z_h$. Second, the additional logarithmic divergence $\ln Q$ arises directly from the charged limit as $Q \to \infty$, where both the island and the turning point approach the horizon $z_p \to z_h$, $z_s \to z_h$.

In order to obtain a dynamical Page curve \cite{Ling:2020laa,Geng:2020fxl,Liu:2022pan}, one must ensure that
\begin{equation}\label{eq:x0Cons}
    \left.\textbf{A}(\hat{\gamma}_c)\right|_{t=0}-\textbf{A}(\hat{\gamma}_d)<0\implies   1 < \frac{z_*(\theta)}{\sqrt{z_h^2 - z_*(\theta)^2}} \cot\frac{\theta}{2}, 
\end{equation}
which finally imposes a constraint on $x_0$.

\begin{figure}
  \centering
    \subfigure[]{\label{fig:saTpTheta}
  \includegraphics[height=0.45\linewidth]{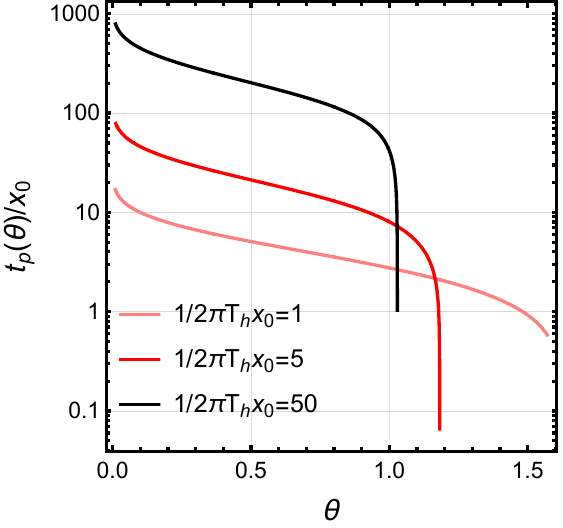}}
    \hspace{0pt}
\subfigure[]{\label{fig:saAt}
  \includegraphics[height=0.45\linewidth]{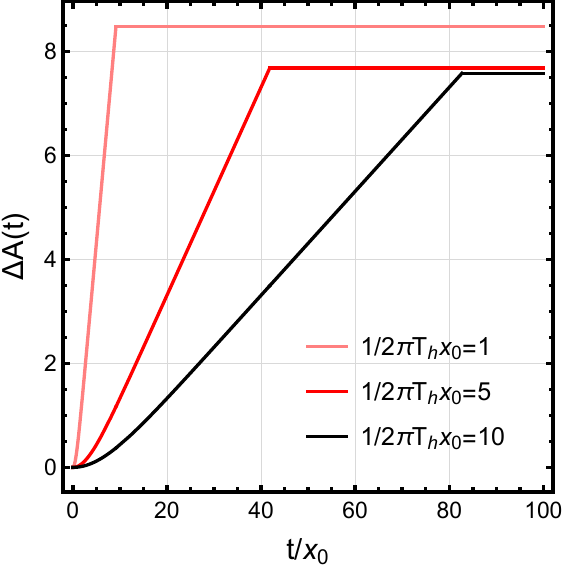}}
\caption{(a): The normalized Page time $t_p(\theta)/x_0$ as a function of the angle $\theta$. (b): Time evolution of the subtracted area $\Delta \textbf{A}(t)$ versus the normalized time $t/x_0$, emphasizing the minimization prescription $\Delta \textbf{A}(t):=\min[\textbf{A}(\hat{\gamma_c})|_{t}, \textbf{A}(\hat{\gamma_d})]-\textbf{A}(\hat{\gamma_c})|_{t=0}$. The electric charge of the black hole is fixed to be $Q\approx10.788$.}
\end{figure}

Assuming this condition is satisfied, the Page time $t_p$ is then determined by 
\begin{equation}\label{eq:pagetime}
    \left.\textbf{A}(\hat{\gamma}_c)\right|_{t=t_p}=\textbf{A}(\hat{\gamma}_d) \implies   t_p = z_h \cosh^{-1}\left( \frac{z_*(\theta)}{\sqrt{z_h^2 - z_*(\theta)^2}} \cot\frac{\theta}{2} \right).
\end{equation}
The Page time is defined as the moment when the areas of the connected and the disconnected surface coincide. It is observed that $t_p$ increases with the electric charge $Q$ (equivalently, decreases with the angle $\theta$). For sufficiently low temperatures, a dynamical Page curve emerges only at large $Q$ -- Fig.~\ref{fig:saTpTheta}.

Again, we now analyze the Page time in two limiting regimes.
First, in the neutral limit $\theta\to \pi/2$, one finds
\begin{equation}
    t_p\simeq z_h\cosh^{-1}\left(\sinh \frac{x_0}{z_h}\right),
\end{equation}
with the constraint being $\sinh(x_0/z_h)>1$. The existence of a lower bound for $x_0$ for obtaining a dynamical Page curve indicates that, in this limit, the brane system itself at $(x\leq 0)$ does not carry sufficient DOF to be entangled with the bath, which is consistent with \cite{Ling:2020laa}. Furthermore, for sufficiently large $x_0\gg z_h$, the above expression reduces to $$t_p \simeq x_0,$$ which implies that the Page time coincides with the time required of the entanglement to propagate from the brane black hole at $x=0$ to the depth of the bath at $x=x_0$ \cite{Liu:2013iza,Liu:2013qca,Leichenauer:2015xra}.

Second, in the charged limit $\theta \ll 1$, one obtains
\begin{equation}
    t_p \simeq x_0 + 3 z_h \ln\frac{2}{\theta}.
\end{equation}
In this regime, the constraint on $x_0$ is always satisfied, since the RHS of \eqref{eq:x0Cons} behaves as $$\text{RHS}\sim \frac{4 e^{x_0/z_h}}{\theta^3}\gg1.$$
Thus, a dynamical Page curve exists for any non-negative $x_0$. We also observe a parametrically large delay in the Page time, $2\pi T_h\,t_p \sim \ln (1/\theta)$, which implies that the large number of DOF on the brane requires an extremely long time to be radiated away via Hawking emission.

Finally, we investigate the time evolution of the holographic entanglement entropy, known as the Page curve, as illustrated in Fig.~\ref{fig:saAt}. The dynamics of the subtracted area $\Delta \mathbf{A}(t)$ reveal two distinct growth regimes before saturation:
\begin{itemize}
    \item \textbf{Early-time Quadratic Growth:} At the early stage ($2\pi T_h\, t\ll 1$), the entanglement entropy exhibits a universal quadratic growth \eqref{eq:earlygrowth}. This regime corresponds to the pre-thermalization phase, where the exchange of Hawking modes between the brane and bath has just commenced, and local thermal equilibrium has not yet been established.
    \item \textbf{Late-time Linear Growth:} As time progresses ($2\pi T_h\, t\gg 1$), the accumulation of the Hawking modes reaches a stable rate determined by the Hawking temperature \eqref{eq:lategrowth}, reflecting the ballistic spreading of entanglement across the system.
\end{itemize}
Intriguingly, the saturation entropy is found to increase with temperature, suggesting that the DOF on the brane are also positively correlated with the black hole temperature. Conversely, the Page time is found to decrease with temperature, indicating that the rate of entanglement exchange is thermally accelerated.
\subsubsection{2d MS black holes on the brane}
\begin{figure}
  \centering
\subfigure[]{\label{fig:snZsZp}
  \includegraphics[height=0.38\linewidth]{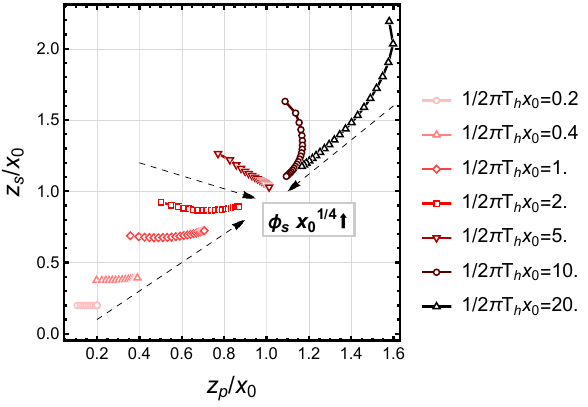}}
    \hspace{0pt}
      \subfigure[]{\label{fig:snAPhis}
  \includegraphics[height=0.38\linewidth]{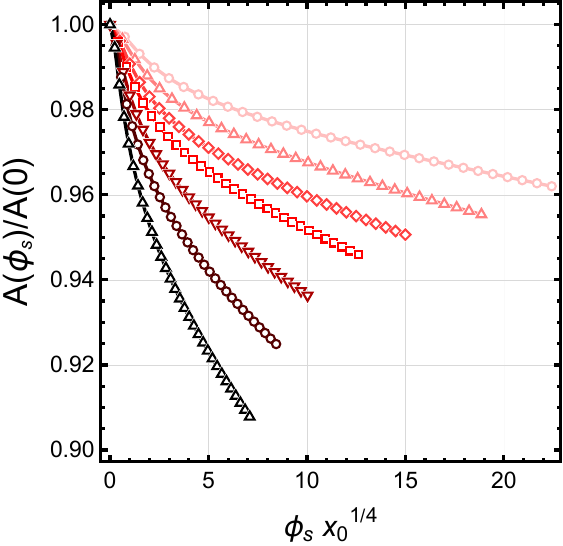}}
\caption{The numerical solutions with $\phi_s\neq0$ of (a): the turning point ratio $z_*(\phi_s)$ as a function of the ratio of the island $z_p$,  and (b): the area ratio of the disconnected extremal surface $\textbf{A}(\hat{\gamma_d})$ normalized by their value at $\phi_s=0$, for varying Hawking temperatures $1/2\pi T_hx_0$, but fixed electric charge $Q\approx2.632$.}
\end{figure}

Activating the scalar field $\Phi$, we transition from the bare Maxwell background to the MS black hole on the brane. Due to the backreaction of the scalar hair, the brane profile and the corresponding transversality conditions can no longer be decoupled analytically and must be solved via our numerical scheme.
 
Turning on the dimensionless scalar source $\phi_s x_0^{1/4}$, the geometric configuration of the extremal surface $\hat{\gamma}_d$ is significantly modified. As illustrated in Fig.~\ref{fig:snZsZp}, for a given temperature, increasing the scalar source (indicated by the dashed arrows) strongly deforms the embedding of $\hat{\gamma}_d$. Interestingly, this deformation exhibits a distinct temperature-dependent susceptibility. In the low-temperature regime (darker curves, where no dynamical Page curve exists), the geometric configuration is highly sensitive to the scalar hair, being rapidly driven towards a strongly deformed state. Conversely, in the high-temperature regime (lighter pink curves, which admit dynamical Page curves), the extremal surface remains rigid.

This phenomenon originates from the competition between the thermal effect and the scalar hair on the brane. At high temperatures, the extremal surface $\hat{\gamma}_d$ firmly hugs the horizon to minimize its area, creating a highly rigid configuration where the scalar hair acts merely as a sub-dominant perturbation. However, at low temperatures, the horizon retreats deep into the bulk. Consequently, the scalar hair easily takes over as the dominant influence, efficiently altering the extremal surface's embedding.

This geometric deformation directly impacts the entanglement entropy -- Fig.~\ref{fig:snAPhis}. We observe a decrease in the normalized area with increasing $\phi_s$. As mentioned in \eqref{eq:effmass}, this suppression indicates that the scalar hair effectively reduces the DOF available to be entangled with the bath. Notably, this suppression exhibits a strong temperature dependence. At higher temperatures, the depletion of DOF by the scalar hair is largely masked by the robust thermal excitations, leading to a mild suppression. Conversely, in the low-temperature regime, thermal excitations are largely frozen out, and the system approaches its vacuum state, rendering a severe suppression.

\begin{figure}
  \centering
    \subfigure[]{\label{fig:snTpPhis}
  \includegraphics[height=0.45\linewidth]{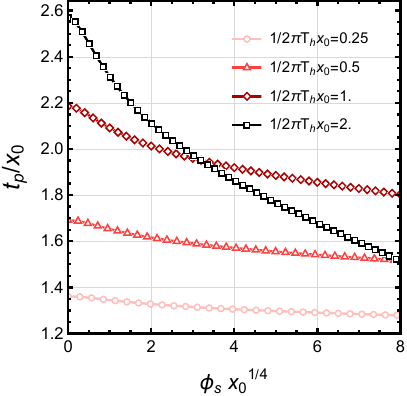}}
    \hspace{0pt}
\subfigure[]{\label{fig:snAt}
  \includegraphics[height=0.45\linewidth]{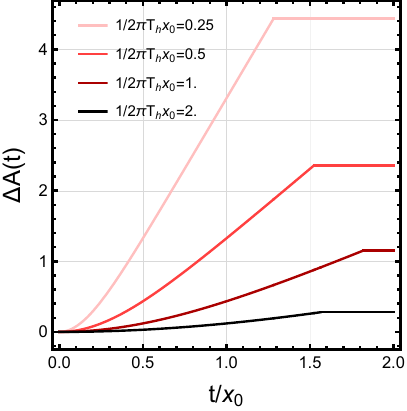}}
\caption{(a): The normalized Page time $t_p(\phi_s)/x_0$ as a function of the normalized scalar source $\phi_s$. (b): Time evolution of the subtracted area $\Delta \textbf{A}(t)$ versus the normalized time $t/x_0$, emphasizing the minimization prescription $\Delta \textbf{A}(t):=\min[\textbf{A}(\hat{\gamma_c})|_{t}, \textbf{A}(\hat{\gamma_d})]-\textbf{A}(\hat{\gamma_c})|_{t=0}$. In both subfigures, the electric charge of the black hole is fixed to be $Q\approx2.632$, while for (b), the normalized scalar source is fixed to be $\phi_s x_0^{1/4}=7.3$.}
\end{figure}

Consequently, the normalized Page time $t_p/x_0$ decreases as the scalar source -- Fig.~\ref{fig:snTpPhis}. For a fixed temperature, this reduction in Page time does not imply that the scalar field accelerates the rate of entanglement exchange \eqref{eq:confunctional-h}. Instead, the earlier onset of the Page time is a direct consequence of a lowered saturation ceiling $\textbf{A}(\hat{\gamma}_d)$. Because the scalar hair significantly suppresses the total entanglement, the steadily growing connected area intercepts this reduced threshold much earlier.

Taken together, the scale at which the Page time emerges is governed by the competition between thermal excitations and the DOF on the brane. On one hand, a higher temperature excites more Hawking modes for exchange per unit time, thereby accelerating the onset of the Page time -- Fig.~\ref{fig:saAt}. On the other hand, if the brane provides more abundant DOF for exchange, the Page time will be delayed further -- Fig.~\ref{fig:saTpTheta} and \ref{fig:snTpPhis}.
Therefore, due to the lack of sufficient thermal excitations at low temperatures, the influence of the DOF on the Page time will become dominant --- reducing the charge or injecting a larger scalar source will significantly alter the scale at which the Page time emerges. This discrepancy creates a ``crossing point'' in -- Fig.~\ref{fig:saTpTheta} and \ref{fig:snTpPhis}, beyond which the Page time of the low-temperature system becomes shorter than its high-temperature counterpart.

Finally, we exhibit the full time evolutions of the subtracted area $\Delta \textbf{A}(t)$ for the MS black hole -- Fig.~\ref{fig:snAt}. Despite the non-trivial backreaction of the scalar field, the universal dynamical features of the Page curve remain robust before saturation:
At the initial stage ($t \ll z_h$), the entanglement entropy still exhibits the universal quadratic growth $\Delta \textbf{A} \propto t^2$, corresponding to the pre-thermalization phase where local equilibrium has not yet been established. As time progresses, the curve transitions into a stable linear growth regime, reflecting the ballistic spreading of Hawking modes across the system.

\section{Conclusions and discussions}\label{sec:3}
In this work, we have systematically investigated the entanglement dynamics and the Page curve of a two-dimensional MS black hole coupled to flat baths, utilizing the framework of double holography. By evaluating the holographic entanglement entropy via the island formula, we provided a comprehensive picture of how scalar hair backreaction influences the entanglement exchanging process.

First of all, from the bulk gravity perspective, the doubly holographic model has been constructed numerically, where the radiations are dual to a higher dimensional bulk and the MS black hole  serves as the Planck brane. 
In two dimensions, the bulk geometry is fixed to be BTZ black holes, while the static profiles of the brane MS black hole are numerically solved via Chebyshev-Lobatto collocations and Newton-Raphson iteration. We have found that scalar hair can provide a negative potential for the system through a negative effective mass squared  ($M^2 < 0$), thereby reducing the local tension near the horizon. This will further reduce the  DOF on the brane that can be entangled with the bath.

The investigation of the entanglement entropy was carried out through both analytical and numerical schemes. In both scenarios, the growth rate of the entropy exhibits a universal behavior --- at early times, the entropy exhibits a  quadratic growth, indicating the exchange of Hawking modes between the brane and bath has just commenced, and local thermal equilibrium has not yet been
established; while at late times, the entropy exhibits a linear growth, indicating the exchanging process reaches a stable rate at a velocity $v_E=2\pi T_h$.

In the bare Maxwell cases ($\phi = 0$), the expressions of the Page time and the Page curve can be obtained analytically. Explicitly, we found that in the neutral limit ($\theta \to \pi/2$), a dynamical Page curve emerges only when the brane subsystem exceeds a critical size, ensuring sufficient DOF for entanglement, which is consistent with the literature \cite{Ling:2020laa,Geng:2020qvw,Liu:2022pan,Liu:2024cmv}. Conversely, in the charged limit ($\theta \to 0$), the extensive DOF on the brane lead to a parametrically delayed Page time, scaling as $t_p \sim \ln(1/\theta)$.

In the MS cases ($\phi > 0$), these counterparts can only be determined numerically, due to the backreaction from the scalar hair. We found that the scalar hair significantly shortens the Page time solely by compressing the DOF on the brane, while the pre-saturation growth rate remains invariant, determined entirely by the Hawking temperature. Moreover, we uncovered a competition between the thermal excitation and the scalar hair: the high-temperature system is protected by robust thermal excitations, and thus is less affected by scalar hair, while the low-temperature system is exactly the opposite. This competition generates a crossing point in the parameter space, inverting the hierarchy of Page times observed in hairless black holes.

From the boundary perspective, where the brane dynamics are holographically dual to a lower-dimensional quantum system coupled to a thermal bath \cite{Almheiri:2019hni}, injecting a scalar source corresponds to an explicit symmetry breaking (distinct from the spontaneous case \cite{Hartnoll:2008vx}). In standard holographic superconductors governed by spontaneous breaking, the emergence of the scalar condensation lowers the entanglement entropy relative to the normal phase \cite{Albash:2012pd,Fan:2013wba,Kuang:2014kha,Garcia-Garcia:2015emb,Wang:2023nes}. While the normal phase possesses more active DOF available to generate entanglement, these DOF are locked into the condensate in the superconducting phase. Although our model utilizes an explicit source directly injected on the brane instead of a bulk thermal trigger, the underlying physical mechanism remains identical: the resulting operator condensate acts as a localized constraint that locks the active DOF, systematically suppressing the entanglement capacity of the system.

It is also intriguing to compare the Page curve of the charged black hole on the brane with that of charged black hole in the bulk \cite{Ling:2020laa}. In the scenario of a charged bulk black hole, increasing the charge source towards extremality intrinsically lowers the Hawking temperature. This thermodynamic suppression not only freezes the exchanging rate of Hawking modes, but also reduces the total entanglement. On the contrary, our brane black hole is in eternal equilibrium with the boundary thermal baths. Therefore, its temperature is forced to remain invariant upon the increase of the charge source. This decoupling ensures a constant rate of entanglement exchange, and an enhancement of the total entanglement.

This work can be extended in two natural directions:
\begin{itemize}
\item \textbf{Nonlinear Evolutions on the Brane} The current framework relies on a static geometry on the brane. A natural extension is to generalize this setup to a dynamically evolving brane, such as modeling a gravitational collapse on the brane. It is crucial to emphasize that in such a scenario, the classical fields undergo non-equilibrium evolution, while the quantum matter sector remains in thermal equilibrium with the bath. Investigating the Page curve in this context would provide a opportunity to explore the intricate fusion of classical gravitational and quantum entanglement dynamics.

\item \textbf{Higher-Dimensional Generalizations} Extending the current $2$-dimensional brane black hole to higher dimensions (e.g., a $4$-dimensional brane embedded in a $5$-dimensional bulk) presents both physical richness and significant technical challenges. Unlike the present setup governed by ordinary differential equations, gravitational DOF in a higher-dimensional framework would propagate into the bulk and break the symmetry, rendering the bulk geometry and the brane embedding highly inhomogeneous. Solving this coupled system involves tackling complicated partial differential equations via the Einstein-DeTurck formulation. We leave these intriguing generalizations for future investigations.

\end{itemize}

\section*{Acknowledgments}
We are grateful to Sun Yuan, Zhu Qinghua for the helpful discussions. Liu Yuxuan special thanks to Cao Peiwen and Liu Ziven for supporting his work. LYX is supported by the Natural Science Foundation of China under Grant No.~12405079, the Natural Science Foundation of Hunan Province, China (Grant No.~2025JJ60062), and Research start-up funds from the Central South University.
\bibliographystyle{unsrt}

\bibliography{refs}

@article{hawking1974black,
  title={Black hole explosions?},
  author={Hawking, Stephen W},
  journal={Nature},
  volume={248},
  number={5443},
  pages={30--31},
  year={1974},
  publisher={Nature Publishing Group}
}

@incollection{hawking1975particle,
  title={Particle creation by black holes},
  author={Hawking, Stephen W},
  booktitle={Euclidean quantum gravity},
  pages={167--188},
  year={1975},
  publisher={World Scientific}
}

@article{hawking1976breakdown,
  title={Breakdown of predictability in gravitational collapse},
  author={Hawking, Stephen W},
  journal={Physical Review D},
  volume={14},
  number={10},
  pages={2460},
  year={1976},
  publisher={APS}
}

@article{Page:1993wv,
    author = "Page, Don N.",
    title = "{Information in black hole radiation}",
    eprint = "hep-th/9306083",
    archivePrefix = "arXiv",
    reportNumber = "ALBERTA-THY-24-93",
    doi = "10.1103/PhysRevLett.71.3743",
    journal = "Phys. Rev. Lett.",
    volume = "71",
    pages = "3743--3746",
    year = "1993"
}

@article{Maldacena:2001kr,
    author = "Maldacena, Juan Martin",
    title = "{Eternal black holes in anti-de Sitter}",
    eprint = "hep-th/0106112",
    archivePrefix = "arXiv",
    reportNumber = "NSF-ITP-01-59",
    doi = "10.1088/1126-6708/2003/04/021",
    journal = "JHEP",
    volume = "04",
    pages = "021",
    year = "2003"
}

@article{Page:2004xp,
    author = "Page, Don N.",
    title = "{Hawking radiation and black hole thermodynamics}",
    eprint = "hep-th/0409024",
    archivePrefix = "arXiv",
    reportNumber = "ALBERTA-THY-18-04",
    doi = "10.1088/1367-2630/7/1/203",
    journal = "New J. Phys.",
    volume = "7",
    pages = "203",
    year = "2005"
}

@article{Takayanagi:2011zk,
    author = "Takayanagi, Tadashi",
    title = "{Holographic Dual of BCFT}",
    eprint = "1105.5165",
    archivePrefix = "arXiv",
    primaryClass = "hep-th",
    reportNumber = "IPMU11-0091",
    doi = "10.1103/PhysRevLett.107.101602",
    journal = "Phys. Rev. Lett.",
    volume = "107",
    pages = "101602",
    year = "2011"
}

@article{Nozaki:2012qd,
    author = "Nozaki, Masahiro and Takayanagi, Tadashi and Ugajin, Tomonori",
    title = "{Central Charges for BCFTs and Holography}",
    eprint = "1205.1573",
    archivePrefix = "arXiv",
    primaryClass = "hep-th",
    reportNumber = "YITP-12-42, IPMU12-0087",
    doi = "10.1007/JHEP06(2012)066",
    journal = "JHEP",
    volume = "06",
    pages = "066",
    year = "2012"
}

@article{Page:2013dx,
    author = "Page, Don N.",
    title = "{Time Dependence of Hawking Radiation Entropy}",
    eprint = "1301.4995",
    archivePrefix = "arXiv",
    primaryClass = "hep-th",
    doi = "10.1088/1475-7516/2013/09/028",
    journal = "JCAP",
    volume = "09",
    pages = "028",
    year = "2013"
}

@article{Hartman:2013qma,
    author = "Hartman, Thomas and Maldacena, Juan",
    title = "{Time Evolution of Entanglement Entropy from Black Hole Interiors}",
    eprint = "1303.1080",
    archivePrefix = "arXiv",
    primaryClass = "hep-th",
    doi = "10.1007/JHEP05(2013)014",
    journal = "JHEP",
    volume = "05",
    pages = "014",
    year = "2013"
}

@article{Lewkowycz:2013nqa,
    author = "Lewkowycz, Aitor and Maldacena, Juan",
    title = "{Generalized gravitational entropy}",
    eprint = "1304.4926",
    archivePrefix = "arXiv",
    primaryClass = "hep-th",
    doi = "10.1007/JHEP08(2013)090",
    journal = "JHEP",
    volume = "08",
    pages = "090",
    year = "2013"
}

@article{Maldacena:2013xja,
    author = "Maldacena, Juan and Susskind, Leonard",
    title = "{Cool horizons for entangled black holes}",
    eprint = "1306.0533",
    archivePrefix = "arXiv",
    primaryClass = "hep-th",
    doi = "10.1002/prop.201300020",
    journal = "Fortsch. Phys.",
    volume = "61",
    pages = "781--811",
    year = "2013"
}

@article{Engelhardt:2014gca,
    author = "Engelhardt, Netta and Wall, Aron C.",
    title = "{Quantum Extremal Surfaces: Holographic Entanglement Entropy beyond the Classical Regime}",
    eprint = "1408.3203",
    archivePrefix = "arXiv",
    primaryClass = "hep-th",
    doi = "10.1007/JHEP01(2015)073",
    journal = "JHEP",
    volume = "01",
    pages = "073",
    year = "2015"
}

@article{penington2020entanglement,
  title={Entanglement wedge reconstruction and the information paradox},
  author={Penington, Geoffrey},
  journal={Journal of High Energy Physics},
  volume={2020},
  number={9},
  pages={1--84},
  year={2020},
  publisher={Springer}
}

@article{Almheiri:2019hni,
    author = "Almheiri, Ahmed and Mahajan, Raghu and Maldacena, Juan and Zhao, Ying",
    title = "{The Page curve of Hawking radiation from semiclassical geometry}",
    eprint = "1908.10996",
    archivePrefix = "arXiv",
    primaryClass = "hep-th",
    doi = "10.1007/JHEP03(2020)149",
    journal = "JHEP",
    volume = "03",
    pages = "149",
    year = "2020"
}

@article{Almheiri:2019psf,
    author = "Almheiri, Ahmed and Engelhardt, Netta and Marolf, Donald and Maxfield, Henry",
    title = "{The entropy of bulk quantum fields and the entanglement wedge of an evaporating black hole}",
    eprint = "1905.08762",
    archivePrefix = "arXiv",
    primaryClass = "hep-th",
    doi = "10.1007/JHEP12(2019)063",
    journal = "JHEP",
    volume = "12",
    pages = "063",
    year = "2019"
}

@article{Chen:2019uhq,
    author = "Chen, Hong Zhe and Fisher, Zachary and Hernandez, Juan and Myers, Robert C. and Ruan, Shan-Ming",
    title = "{Information Flow in Black Hole Evaporation}",
    eprint = "1911.03402",
    archivePrefix = "arXiv",
    primaryClass = "hep-th",
    doi = "10.1007/JHEP03(2020)152",
    journal = "JHEP",
    volume = "03",
    pages = "152",
    year = "2020"
}

@article{Almheiri:2019yqk,
    author = "Almheiri, Ahmed and Mahajan, Raghu and Maldacena, Juan",
    title = "{Islands outside the horizon}",
    journal = "arXiv:1910.11077",
    archivePrefix = "arXiv",
    primaryClass = "hep-th",
    month = "10",
    year = "2019"
}

@article{Penington:2019kki,
    author = "Penington, Geoff and Shenker, Stephen H. and Stanford, Douglas and Yang, Zhenbin",
    title = "{Replica wormholes and the black hole interior}",
    eprint = "1911.11977",
    archivePrefix = "arXiv",
    primaryClass = "hep-th",
    doi = "10.1007/JHEP03(2022)205",
    journal = "JHEP",
    volume = "03",
    pages = "205",
    year = "2022"
}

@article{Almheiri:2019qdq,
    author = "Almheiri, Ahmed and Hartman, Thomas and Maldacena, Juan and Shaghoulian, Edgar and Tajdini, Amirhossein",
    title = "{Replica Wormholes and the Entropy of Hawking Radiation}",
    eprint = "1911.12333",
    archivePrefix = "arXiv",
    primaryClass = "hep-th",
    doi = "10.1007/JHEP05(2020)013",
    journal = "JHEP",
    volume = "05",
    pages = "013",
    year = "2020"
}

@article{Almheiri:2019psy,
    author = "Almheiri, Ahmed and Mahajan, Raghu and Santos, Jorge E.",
    title = "{Entanglement islands in higher dimensions}",
    eprint = "1911.09666",
    archivePrefix = "arXiv",
    primaryClass = "hep-th",
    doi = "10.21468/SciPostPhys.9.1.001",
    journal = "SciPost Phys.",
    volume = "9",
    number = "1",
    pages = "001",
    year = "2020"
}

@article{Rozali:2019day,
    author = "Rozali, Moshe and Sully, James and Van Raamsdonk, Mark and Waddell, Christopher and Wakeham, David",
    title = "{Information radiation in BCFT models of black holes}",
    eprint = "1910.12836",
    archivePrefix = "arXiv",
    primaryClass = "hep-th",
    doi = "10.1007/JHEP05(2020)004",
    journal = "JHEP",
    volume = "05",
    pages = "004",
    year = "2020"
}

@article{Geng:2020qvw,
    author = "Geng, Hao and Karch, Andreas",
    title = "{Massive islands}",
    eprint = "2006.02438",
    archivePrefix = "arXiv",
    primaryClass = "hep-th",
    doi = "10.1007/JHEP09(2020)121",
    journal = "JHEP",
    volume = "09",
    pages = "121",
    year = "2020"
}

@article{Chen:2020uac,
    author = "Chen, Hong Zhe and Myers, Robert C. and Neuenfeld, Dominik and Reyes, Ignacio A. and Sandor, Joshua",
    title = "{Quantum Extremal Islands Made Easy, Part I: Entanglement on the Brane}",
    eprint = "2006.04851",
    archivePrefix = "arXiv",
    primaryClass = "hep-th",
    doi = "10.1007/JHEP10(2020)166",
    journal = "JHEP",
    volume = "10",
    pages = "166",
    year = "2020"
}

@article{Chen:2020hmv,
    author = "Chen, Hong Zhe and Myers, Robert C. and Neuenfeld, Dominik and Reyes, Ignacio A. and Sandor, Joshua",
    title = "{Quantum Extremal Islands Made Easy, Part II: Black Holes on the Brane}",
    eprint = "2010.00018",
    archivePrefix = "arXiv",
    primaryClass = "hep-th",
    doi = "10.1007/JHEP12(2020)025",
    journal = "JHEP",
    volume = "12",
    pages = "025",
    year = "2020"
}

@article{Hernandez:2020nem,
    author = "Hernandez, Juan and Myers, Robert C. and Ruan, Shan-Ming",
    title = "{Quantum extremal islands made easy. Part III. Complexity on the brane}",
    eprint = "2010.16398",
    archivePrefix = "arXiv",
    primaryClass = "hep-th",
    doi = "10.1007/JHEP02(2021)173",
    journal = "JHEP",
    volume = "02",
    pages = "173",
    year = "2021"
}

@article{Alishahiha:2020qza,
    author = "Alishahiha, Mohsen and Faraji Astaneh, Amin and Naseh, Ali",
    title = "{Island in the presence of higher derivative terms}",
    eprint = "2005.08715",
    archivePrefix = "arXiv",
    primaryClass = "hep-th",
    doi = "10.1007/JHEP02(2021)035",
    journal = "JHEP",
    volume = "02",
    pages = "035",
    year = "2021"
}

@article{Hashimoto:2020cas,
    author = "Hashimoto, Koji and Iizuka, Norihiro and Matsuo, Yoshinori",
    title = "{Islands in Schwarzschild black holes}",
    eprint = "2004.05863",
    archivePrefix = "arXiv",
    primaryClass = "hep-th",
    reportNumber = "OU-HET-1053",
    doi = "10.1007/JHEP06(2020)085",
    journal = "JHEP",
    volume = "06",
    pages = "085",
    year = "2020"
}

@article{Almheiri:2020cfm,
    author = "Almheiri, Ahmed and Hartman, Thomas and Maldacena, Juan and Shaghoulian, Edgar and Tajdini, Amirhossein",
    title = "{The entropy of Hawking radiation}",
    eprint = "2006.06872",
    archivePrefix = "arXiv",
    primaryClass = "hep-th",
    doi = "10.1103/RevModPhys.93.035002",
    journal = "Rev. Mod. Phys.",
    volume = "93",
    number = "3",
    pages = "035002",
    year = "2021"
}

@article{Geng:2020fxl,
    author = "Geng, Hao and Karch, Andreas and Perez-Pardavila, Carlos and Raju, Suvrat and Randall, Lisa and Riojas, Marcos and Shashi, Sanjit",
    title = "{Information Transfer with a Gravitating Bath}",
    eprint = "2012.04671",
    archivePrefix = "arXiv",
    primaryClass = "hep-th",
    doi = "10.21468/SciPostPhys.10.5.103",
    journal = "SciPost Phys.",
    volume = "10",
    number = "5",
    pages = "103",
    year = "2021"
}

@article{Ling:2020laa,
    author = "Ling, Yi and Liu, Yuxuan and Xian, Zhuo-Yu",
    title = "{Island in Charged Black Holes}",
    eprint = "2010.00037",
    archivePrefix = "arXiv",
    primaryClass = "hep-th",
    doi = "10.1007/JHEP03(2021)251",
    journal = "JHEP",
    volume = "03",
    pages = "251",
    year = "2021"
}

@article{Anegawa:2020ezn,
    author = "Anegawa, Takanori and Iizuka, Norihiro",
    title = "{Notes on islands in asymptotically flat 2d dilaton black holes}",
    eprint = "2004.01601",
    archivePrefix = "arXiv",
    primaryClass = "hep-th",
    reportNumber = "OU-HET-1051",
    doi = "10.1007/JHEP07(2020)036",
    journal = "JHEP",
    volume = "07",
    pages = "036",
    year = "2020"
}

@article{Hartman:2020swn,
    author = "Hartman, Thomas and Shaghoulian, Edgar and Strominger, Andrew",
    title = "{Islands in Asymptotically Flat 2D Gravity}",
    eprint = "2004.13857",
    archivePrefix = "arXiv",
    primaryClass = "hep-th",
    doi = "10.1007/JHEP07(2020)022",
    journal = "JHEP",
    volume = "07",
    pages = "022",
    year = "2020"
}

@article{Caceres:2020jcn,
    author = "Caceres, Elena and Kundu, Arnab and Patra, Ayan K. and Shashi, Sanjit",
    title = "{Warped information and entanglement islands in AdS/WCFT}",
    eprint = "2012.05425",
    archivePrefix = "arXiv",
    primaryClass = "hep-th",
    reportNumber = "UTTG-23-2020",
    doi = "10.1007/JHEP07(2021)004",
    journal = "JHEP",
    volume = "07",
    pages = "004",
    year = "2021"
}

@article{Li:2020ceg,
    author = "Li, Tianyi and Chu, Jinwei and Zhou, Yang",
    title = "{Reflected Entropy for an Evaporating Black Hole}",
    eprint = "2006.10846",
    archivePrefix = "arXiv",
    primaryClass = "hep-th",
    doi = "10.1007/JHEP11(2020)155",
    journal = "JHEP",
    volume = "11",
    pages = "155",
    year = "2020"
}

@article{Gautason:2020tmk,
    author = "Gautason, F. F. and Schneiderbauer, Lukas and Sybesma, Watse and Thorlacius, L\'arus",
    title = "{Page Curve for an Evaporating Black Hole}",
    eprint = "2004.00598",
    archivePrefix = "arXiv",
    primaryClass = "hep-th",
    doi = "10.1007/JHEP05(2020)091",
    journal = "JHEP",
    volume = "05",
    pages = "091",
    year = "2020"
}

@article{Krishnan:2020oun,
    author = "Krishnan, Chethan and Patil, Vaishnavi and Pereira, Jude",
    title = "{Page Curve and the Information Paradox in Flat Space}",
    eprint = "2005.02993",
    archivePrefix = "arXiv",
    primaryClass = "hep-th",
    month = "5",
    year = "2020"
}

@article{Karlsson:2020uga,
    author = "Karlsson, Anna",
    title = "{Replica wormhole and island incompatibility with monogamy of entanglement}",
    eprint = "2007.10523",
    archivePrefix = "arXiv",
    primaryClass = "hep-th",
    month = "7",
    year = "2020"
}

@article{Marolf:2020xie,
    author = "Marolf, Donald and Maxfield, Henry",
    title = "{Transcending the ensemble: baby universes, spacetime wormholes, and the order and disorder of black hole information}",
    eprint = "2002.08950",
    archivePrefix = "arXiv",
    primaryClass = "hep-th",
    doi = "10.1007/JHEP08(2020)044",
    journal = "JHEP",
    volume = "08",
    pages = "044",
    year = "2020"
}

@article{Balasubramanian:2020jhl,
    author = "Balasubramanian, Vijay and Kar, Arjun and Ross, Simon F. and Ugajin, Tomonori",
    title = "{Spin structures and baby universes}",
    eprint = "2007.04333",
    archivePrefix = "arXiv",
    primaryClass = "hep-th",
    doi = "10.1007/JHEP09(2020)192",
    journal = "JHEP",
    volume = "09",
    pages = "192",
    year = "2020"
}

@article{Chen:2020jvn,
    author = "Chen, Hong Zhe and Fisher, Zachary and Hernandez, Juan and Myers, Robert C. and Ruan, Shan-Ming",
    title = "{Evaporating Black Holes Coupled to a Thermal Bath}",
    eprint = "2007.11658",
    archivePrefix = "arXiv",
    primaryClass = "hep-th",
    doi = "10.1007/JHEP01(2021)065",
    journal = "JHEP",
    volume = "01",
    pages = "065",
    year = "2021"
}

@article{Balasubramanian:2020xqf,
    author = "Balasubramanian, Vijay and Kar, Arjun and Ugajin, Tomonori",
    title = "{Islands in de Sitter space}",
    eprint = "2008.05275",
    archivePrefix = "arXiv",
    primaryClass = "hep-th",
    doi = "10.1007/JHEP02(2021)072",
    journal = "JHEP",
    volume = "02",
    pages = "072",
    year = "2021"
}

@article{Bhattacharya:2020uun,
    author = "Bhattacharya, Aranya and Chanda, Anindya and Maulik, Sabyasachi and Northe, Christian and Roy, Shibaji",
    title = "{Topological shadows and complexity of islands in multiboundary wormholes}",
    eprint = "2010.04134",
    archivePrefix = "arXiv",
    primaryClass = "hep-th",
    doi = "10.1007/JHEP02(2021)152",
    journal = "JHEP",
    volume = "02",
    pages = "152",
    year = "2021"
}

@article{Deng:2020ent,
    author = "Deng, Feiyu and Chu, Jinwei and Zhou, Yang",
    title = "{Defect extremal surface as the holographic counterpart of Island formula}",
    eprint = "2012.07612",
    archivePrefix = "arXiv",
    primaryClass = "hep-th",
    doi = "10.1007/JHEP03(2021)008",
    journal = "JHEP",
    volume = "03",
    pages = "008",
    year = "2021"
}

@article{Sybesma:2020fxg,
    author = "Sybesma, Watse",
    title = "{Pure de Sitter space and the island moving back in time}",
    eprint = "2008.07994",
    archivePrefix = "arXiv",
    primaryClass = "hep-th",
    doi = "10.1088/1361-6382/abff9a",
    journal = "Class. Quant. Grav.",
    volume = "38",
    number = "14",
    pages = "145012",
    year = "2021"
}

@article{Marolf:2020rpm,
    author = "Marolf, Donald and Maxfield, Henry",
    title = "{Observations of Hawking radiation: the Page curve and baby universes}",
    eprint = "2010.06602",
    archivePrefix = "arXiv",
    primaryClass = "hep-th",
    doi = "10.1007/JHEP04(2021)272",
    journal = "JHEP",
    volume = "04",
    pages = "272",
    year = "2021"
}

@article{Balasubramanian:2020coy,
    author = "Balasubramanian, Vijay and Kar, Arjun and Ugajin, Tomonori",
    title = "{Entanglement between two disjoint universes}",
    eprint = "2008.05274",
    archivePrefix = "arXiv",
    primaryClass = "hep-th",
    doi = "10.1007/JHEP02(2021)136",
    journal = "JHEP",
    volume = "02",
    pages = "136",
    year = "2021"
}

@article{KumarBasak:2020ams,
    author = "Kumar Basak, Jaydeep and Basu, Debarshi and Malvimat, Vinay and Parihar, Himanshu and Sengupta, Gautam",
    title = "{Islands for entanglement negativity}",
    eprint = "2012.03983",
    archivePrefix = "arXiv",
    primaryClass = "hep-th",
    doi = "10.21468/SciPostPhys.12.1.003",
    journal = "SciPost Phys.",
    volume = "12",
    number = "1",
    pages = "003",
    year = "2022"
}

@article{Ling:2021vxe,
    author = "Ling, Yi and Liu, Peng and Liu, Yuxuan and Niu, Chao and Xian, Zhuo-Yu and Zhang, Cheng-Yong",
    title = "{Reflected entropy in double holography}",
    eprint = "2109.09243",
    archivePrefix = "arXiv",
    primaryClass = "hep-th",
    doi = "10.1007/JHEP02(2022)037",
    journal = "JHEP",
    volume = "02",
    pages = "037",
    year = "2022"
}

@article{Wang:2021woy,
    author = "Wang, Xuanhua and Li, Ran and Wang, Jin",
    title = {{Islands and Page curves of Reissner-Nordstr\"om black holes}},
    eprint = "2101.06867",
    archivePrefix = "arXiv",
    primaryClass = "hep-th",
    doi = "10.1007/JHEP04(2021)103",
    journal = "JHEP",
    volume = "04",
    pages = "103",
    year = "2021"
}

@article{He:2021mst,
    author = "He, Song and Sun, Yuan and Zhao, Long and Zhang, Yu-Xuan",
    title = "{The universality of islands outside the horizon}",
    eprint = "2110.07598",
    archivePrefix = "arXiv",
    primaryClass = "hep-th",
    doi = "10.1007/JHEP05(2022)047",
    journal = "JHEP",
    volume = "05",
    pages = "047",
    year = "2022"
}

@article{Hollowood:2021lsw,
    author = "Hollowood, Timothy J. and Kumar, S. Prem and Legramandi, Andrea and Talwar, Neil",
    title = "{Grey-body factors, irreversibility and multiple island saddles}",
    eprint = "2111.02248",
    archivePrefix = "arXiv",
    primaryClass = "hep-th",
    doi = "10.1007/JHEP03(2022)110",
    journal = "JHEP",
    volume = "03",
    pages = "110",
    year = "2022"
}

@article{Vardhan:2021mdy,
    author = "Vardhan, Shreya and Kudler-Flam, Jonah and Shapourian, Hassan and Liu, Hong",
    title = "{Mixed-state entanglement and information recovery in thermalized states and evaporating black holes}",
    eprint = "2112.00020",
    archivePrefix = "arXiv",
    primaryClass = "hep-th",
    reportNumber = "MIT-CTP/5362",
    doi = "10.1007/JHEP01(2023)064",
    journal = "JHEP",
    volume = "01",
    pages = "064",
    year = "2023"
}

@article{Kawabata:2021vyo,
    author = "Kawabata, Kohki and Nishioka, Tatsuma and Okuyama, Yoshitaka and Watanabe, Kento",
    title = "{Replica wormholes and capacity of entanglement}",
    eprint = "2105.08396",
    archivePrefix = "arXiv",
    primaryClass = "hep-th",
    reportNumber = "YITP-21-45",
    doi = "10.1007/JHEP10(2021)227",
    journal = "JHEP",
    volume = "10",
    pages = "227",
    year = "2021"
}

@article{Kawabata:2021hac,
    author = "Kawabata, Kohki and Nishioka, Tatsuma and Okuyama, Yoshitaka and Watanabe, Kento",
    title = "{Probing Hawking radiation through capacity of entanglement}",
    eprint = "2102.02425",
    archivePrefix = "arXiv",
    primaryClass = "hep-th",
    reportNumber = "YITP-21-08",
    doi = "10.1007/JHEP05(2021)062",
    journal = "JHEP",
    volume = "05",
    pages = "062",
    year = "2021"
}

@article{Geng:2021iyq,
    author = {Geng, Hao and L\"ust, Severin and Mishra, Rashmish K. and Wakeham, David},
    title = "{Holographic BCFTs and Communicating Black Holes}",
    eprint = "2104.07039",
    archivePrefix = "arXiv",
    primaryClass = "hep-th",
    doi = "10.1007/JHEP08(2021)003",
    journal = "jhep",
    volume = "08",
    pages = "003",
    year = "2021"
}

@article{Geng:2021mic,
    author = "Geng, Hao and Karch, Andreas and Perez-Pardavila, Carlos and Raju, Suvrat and Randall, Lisa and Riojas, Marcos and Shashi, Sanjit",
    title = "{Entanglement phase structure of a holographic BCFT in a black hole background}",
    eprint = "2112.09132",
    archivePrefix = "arXiv",
    primaryClass = "hep-th",
    reportNumber = "UTTG-27-2021",
    doi = "10.1007/JHEP05(2022)153",
    journal = "JHEP",
    volume = "05",
    pages = "153",
    year = "2022"
}

@article{Chou:2021boq,
    author = "Chou, Chia-Jui and Lao, Hans B. and Yang, Yi",
    title = "{Page curve of effective Hawking radiation}",
    eprint = "2111.14551",
    archivePrefix = "arXiv",
    primaryClass = "hep-th",
    doi = "10.1103/PhysRevD.106.066008",
    journal = "Phys. Rev. D",
    volume = "106",
    number = "6",
    pages = "066008",
    year = "2022"
}

@article{Miyata:2021qsm,
    author = "Miyata, Akihiro and Ugajin, Tomonori",
    title = "{Entanglement between two evaporating black holes}",
    eprint = "2111.11688",
    archivePrefix = "arXiv",
    primaryClass = "hep-th",
    reportNumber = "UT-Komaba/21-5",
    doi = "10.1007/JHEP09(2022)009",
    journal = "JHEP",
    volume = "09",
    pages = "009",
    year = "2022"
}

@article{Akal:2021dqt,
    author = "Akal, Ibrahim and Kawamoto, Taishi and Ruan, Shan-Ming and Takayanagi, Tadashi and Wei, Zixia",
    title = "{Page curve under final state projection}",
    eprint = "2112.08433",
    archivePrefix = "arXiv",
    primaryClass = "hep-th",
    reportNumber = "YITP-21-158, IPMU21-0086, YITP-21-158; IPMU21-0086",
    doi = "10.1103/PhysRevD.105.126026",
    journal = "Phys. Rev. D",
    volume = "105",
    number = "12",
    pages = "126026",
    year = "2022"
}

@article{Renner:2021qbe,
    author = "Renner, Renato and Wang, Jinzhao",
    title = "{The black hole information puzzle and the quantum de Finetti theorem}",
    eprint = "2110.14653",
    archivePrefix = "arXiv",
    primaryClass = "hep-th",
    month = "10",
    year = "2021"
}

@article{Balasubramanian:2021wgd,
    author = "Balasubramanian, Vijay and Kar, Arjun and Ugajin, Tomonori",
    title = "{Entanglement between two gravitating universes}",
    eprint = "2104.13383",
    archivePrefix = "arXiv",
    primaryClass = "hep-th",
    reportNumber = "YITP-21-39",
    doi = "10.1088/1361-6382/ac3c8b",
    journal = "Class. Quant. Grav.",
    volume = "39",
    number = "17",
    pages = "174001",
    year = "2022"
}

@article{Bhattacharya:2021nqj,
    author = "Bhattacharya, Aranya and Bhattacharyya, Arpan and Nandy, Pratik and Patra, Ayan K.",
    title = "{Bath deformations, islands, and holographic complexity}",
    eprint = "2112.06967",
    archivePrefix = "arXiv",
    primaryClass = "hep-th",
    doi = "10.1103/PhysRevD.105.066019",
    journal = "Phys. Rev. D",
    volume = "105",
    number = "6",
    pages = "066019",
    year = "2022"
}

@article{Bhattacharya:2021dnd,
    author = "Bhattacharya, Aranya and Bhattacharyya, Arpan and Nandy, Pratik and Patra, Ayan K.",
    title = "{Partial islands and subregion complexity in geometric secret-sharing model}",
    eprint = "2109.07842",
    archivePrefix = "arXiv",
    primaryClass = "hep-th",
    doi = "10.1007/JHEP12(2021)091",
    journal = "JHEP",
    volume = "12",
    pages = "091",
    year = "2021"
}

@article{Caceres:2021fuw,
    author = "Caceres, Elena and Kundu, Arnab and Patra, Ayan K. and Shashi, Sanjit",
    title = "{Page curves and bath deformations}",
    eprint = "2107.00022",
    archivePrefix = "arXiv",
    primaryClass = "hep-th",
    doi = "10.21468/SciPostPhysCore.5.2.033",
    journal = "SciPost Phys. Core",
    volume = "5",
    pages = "033",
    year = "2022"
}

@article{Bhattacharya:2021jrn,
    author = "Bhattacharya, Aranya and Bhattacharyya, Arpan and Nandy, Pratik and Patra, Ayan K.",
    title = "{Islands and complexity of eternal black hole and radiation subsystems for a doubly holographic model}",
    eprint = "2103.15852",
    archivePrefix = "arXiv",
    primaryClass = "hep-th",
    doi = "10.1007/JHEP05(2021)135",
    journal = "JHEP",
    volume = "05",
    pages = "135",
    year = "2021"
}

@article{Peng:2021vhs,
    author = "Peng, Cheng and Tian, Jia and Yu, Jianghui",
    title = "{Baby universes, ensemble averages and factorizations with matters}",
    eprint = "2111.14856",
    archivePrefix = "arXiv",
    primaryClass = "hep-th",
    journal={arXiv preprint arXiv:2111.14856},
    month = "11",
    year = "2021"
}

@article{Miyata:2021ncm,
    author = "Miyata, Akihiro and Ugajin, Tomonori",
    title = "{Evaporation of black holes in flat space entangled with an auxiliary universe}",
    eprint = "2104.00183",
    archivePrefix = "arXiv",
    primaryClass = "hep-th",
    reportNumber = "UT-Komaba/21-3, YITP-21-27",
    doi = "10.1093/ptep/ptab163",
    journal = "PTEP",
    volume = "2022",
    number = "1",
    pages = "013B13",
    year = "2022"
}

@article{Suzuki:2022xwv,
    author = "Suzuki, Kenta and Takayanagi, Tadashi",
    title = "{BCFT and Islands in two dimensions}",
    eprint = "2202.08462",
    archivePrefix = "arXiv",
    primaryClass = "hep-th",
    reportNumber = "YITP-22-14, IPMU22-0002",
    doi = "10.1007/JHEP06(2022)095",
    journal = "JHEP",
    volume = "06",
    pages = "095",
    year = "2022"
}

@article{Suzuki:2022yru,
    author = "Suzuki, Yu-ki and Terashima, Seiji",
    title = "{On the dynamics in the AdS/BCFT correspondence}",
    eprint = "2205.10600",
    archivePrefix = "arXiv",
    primaryClass = "hep-th",
    reportNumber = "YITP-22-50",
    doi = "10.1007/JHEP09(2022)103",
    journal = "JHEP",
    volume = "09",
    pages = "103",
    year = "2022"
}

@article{Afrasiar:2022ebi,
    author = "Afrasiar, Mir and Kumar Basak, Jaydeep and Chandra, Ashish and Sengupta, Gautam",
    title = "{Islands for Entanglement Negativity in Communicating Black Holes}",
    journal = "arXiv:2205.07903",
    archivePrefix = "arXiv",
    primaryClass = "hep-th",
    month = "5",
    year = "2022"
}

@article{Liu:2022pan,
    author = "Liu, Yuxuan and Xian, Zhuo-Yu and Peng, Cheng and Ling, Yi",
    title = "{Addendum to: Black holes entangled by radiation}",
    eprint = "2205.14596",
    archivePrefix = "arXiv",
    primaryClass = "hep-th",
    doi = "10.1007/JHEP11(2022)043",
    journal = "JHEP",
    volume = "11",
    pages = "043",
    year = "2022"
}

@article{Erdmenger:2013dpa,
    author = "Erdmenger, Johanna and Hoyos, Carlos and O'Bannon, Andy and Wu, Jackson",
    title = "{A Holographic Model of the Kondo Effect}",
    eprint = "1310.3271",
    archivePrefix = "arXiv",
    primaryClass = "hep-th",
    reportNumber = "DAMTP-2013-57, MPP-2013-128, OUTP-13-21P, TAUP-2977-13",
    doi = "10.1007/JHEP12(2013)086",
    journal = "JHEP",
    volume = "12",
    pages = "086",
    year = "2013"
}

@article{Erdmenger:2014xya,
    author = "Erdmenger, Johanna and Flory, Mario and Newrzella, Max-Niklas",
    title = "{Bending branes for DCFT in two dimensions}",
    eprint = "1410.7811",
    archivePrefix = "arXiv",
    primaryClass = "hep-th",
    reportNumber = "MPP-2014-372",
    doi = "10.1007/JHEP01(2015)058",
    journal = "JHEP",
    volume = "01",
    pages = "058",
    year = "2015"
}

@article{Erdmenger:2015spo,
    author = "Erdmenger, Johanna and Flory, Mario and Hoyos, Carlos and Newrzella, Max-Niklas and Wu, Jackson M. S.",
    title = "{Entanglement Entropy in a Holographic Kondo Model}",
    eprint = "1511.03666",
    archivePrefix = "arXiv",
    primaryClass = "hep-th",
    reportNumber = "MPP-2015-248, FPAUO-15-15",
    doi = "10.1002/prop.201500099",
    journal = "Fortsch. Phys.",
    volume = "64",
    pages = "109--130",
    year = "2016"
}

@article{Erdmenger:2015xpq,
    author = "Erdmenger, J. and Flory, M. and Hoyos, C. and Newrzella, M-N. and O'Bannon, A. and Wu, J.",
    editor = "Argurio, Riccardo and Bobev, Nikolay and Boulanger, Nicolas and Craps, Ben and Henneaux, Marc and Hertog, Thomas and Sevrin, Alex and Van Proeyen, Antoine and Van Riet, Thomas",
    title = "{Holographic impurities and Kondo effect}",
    eprint = "1511.09362",
    archivePrefix = "arXiv",
    primaryClass = "hep-th",
    reportNumber = "FPAUO-15-17",
    doi = "10.1002/prop.201500079",
    journal = "Fortsch. Phys.",
    volume = "64",
    pages = "322--329",
    year = "2016"
}

@article{Erdmenger:2016msd,
    author = "Erdmenger, Johanna and Flory, Mario and Newrzella, Max-Niklas and Strydom, Migael and Wu, Jackson M. S.",
    title = "{Quantum Quenches in a Holographic Kondo Model}",
    eprint = "1612.06860",
    archivePrefix = "arXiv",
    primaryClass = "hep-th",
    reportNumber = "MPP-2016-332",
    doi = "10.1007/JHEP04(2017)045",
    journal = "JHEP",
    volume = "04",
    pages = "045",
    year = "2017"
}

@article{Erdmenger:2020fqe,
    author = "Erdmenger, Johanna",
    title = "{Holographic Kondo Models}",
    doi = "10.1007/978-3-030-35473-2_6",
    journal = "Springer Proc. Phys.",
    volume = "239",
    pages = "155--194",
    year = "2020"
}

@article{Chandrasekaran:2020qtn,
    author = "Chandrasekaran, Venkatesa and Miyaji, Masamichi and Rath, Pratik",
    title = "{Including contributions from entanglement islands to the reflected entropy}",
    eprint = "2006.10754",
    archivePrefix = "arXiv",
    primaryClass = "hep-th",
    doi = "10.1103/PhysRevD.102.086009",
    journal = "Phys. Rev. D",
    volume = "102",
    number = "8",
    pages = "086009",
    year = "2020"
}

@article{Liu:2024cmv,
    author = "Liu, Yuxuan and Jian, Shao-Kai and Ling, Yi and Xian, Zhuo-Yu",
    title = "{Entanglement inside a black hole before the Page time}",
    eprint = "2401.04706",
    archivePrefix = "arXiv",
    primaryClass = "hep-th",
    month = "1",
    year = "2024"
}

@article{Liu:2023ggg,
    author = "Liu, Yuxuan and Chen, Qian and Ling, Yi and Peng, Cheng and Tian, Yu and Xian, Zhuo-Yu",
    title = "{Addendum to: Entanglement of defect subregions in double holography}",
    eprint = "2312.08025",
    archivePrefix = "arXiv",
    primaryClass = "hep-th",
    doi = "10.1007/JHEP09(2024)194",
    journal = "JHEP",
    volume = "09",
    pages = "194",
    year = "2024"
}

@article{Sully:2020pza,
    author = "Sully, James and Van Raamsdonk, Mark and Wakeham, David",
    title = "{BCFT entanglement entropy at large central charge and the black hole interior}",
    eprint = "2004.13088",
    archivePrefix = "arXiv",
    primaryClass = "hep-th",
    doi = "10.1007/JHEP03(2021)167",
    journal = "JHEP",
    volume = "03",
    pages = "167",
    year = "2021"
}

@article{Liu:2013iza,
    author = "Liu, Hong and Suh, S. Josephine",
    title = "{Entanglement Tsunami: Universal Scaling in Holographic Thermalization}",
    eprint = "1305.7244",
    archivePrefix = "arXiv",
    primaryClass = "hep-th",
    reportNumber = "MIT-CTP/4475, MIT-CTP-4475",
    doi = "10.1103/PhysRevLett.112.011601",
    journal = "Phys. Rev. Lett.",
    volume = "112",
    pages = "011601",
    year = "2014"
}

@article{Liu:2013qca,
    author = "Liu, Hong and Suh, S. Josephine",
    title = "{Entanglement growth during thermalization in holographic systems}",
    eprint = "1311.1200",
    archivePrefix = "arXiv",
    primaryClass = "hep-th",
    reportNumber = "MIT-CTP-4510, MIT-CTP 4510",
    doi = "10.1103/PhysRevD.89.066012",
    journal = "Phys. Rev. D",
    volume = "89",
    number = "6",
    pages = "066012",
    year = "2014"
}

@article{Leichenauer:2015xra,
    author = "Leichenauer, Stefan and Moosa, Mudassir",
    title = "{Entanglement Tsunami in (1+1)-Dimensions}",
    eprint = "1505.04225",
    archivePrefix = "arXiv",
    primaryClass = "hep-th",
    doi = "10.1103/PhysRevD.92.126004",
    journal = "Phys. Rev. D",
    volume = "92",
    pages = "126004",
    year = "2015"
}

@article{Miao:2024olz,
    author = "Miao, Rong-Xin and Xie, Zi-Bing",
    title = "{Holographic entanglement entropy for brane-world higher derivative gravity}",
    eprint = "2410.18314",
    archivePrefix = "arXiv",
    primaryClass = "hep-th",
    doi = "10.1007/JHEP03(2025)015",
    journal = "JHEP",
    volume = "03",
    pages = "015",
    year = "2025"
}

@article{Cui:2023gtf,
    author = "Cui, Zheng-Quan and Guo, Yu and Miao, Rong-Xin",
    title = "{Cone holography with Neumann boundary conditions and brane-localized gauge fields}",
    eprint = "2312.16463",
    archivePrefix = "arXiv",
    primaryClass = "hep-th",
    doi = "10.1007/JHEP03(2024)158",
    journal = "JHEP",
    volume = "03",
    pages = "158",
    year = "2024"
}

@article{Guo:2023fly,
    author = "Guo, Yu and Miao, Rong-Xin",
    title = "{Page curves on codim-m and charged branes}",
    doi = "10.1140/epjc/s10052-023-12026-4",
    journal = "Eur. Phys. J. C",
    volume = "83",
    number = "9",
    pages = "847",
    year = "2023"
}

@article{Miao:2023mui,
    author = "Miao, Rong-Xin",
    title = "{Ghost problem, spectrum identities and various constraints on brane-localized gravity}",
    eprint = "2310.16297",
    archivePrefix = "arXiv",
    primaryClass = "hep-th",
    doi = "10.1007/JHEP06(2024)043",
    journal = "JHEP",
    volume = "06",
    pages = "043",
    year = "2024"
}

@article{Hao:2025ocu,
    author = "Hao, Peng-Xiang and Ogawa, Naoki and Takayanagi, Tadashi and Waki, Takahiro",
    title = "{Flat space holography via AdS/BCFT}",
    eprint = "2509.00652",
    archivePrefix = "arXiv",
    primaryClass = "hep-th",
    reportNumber = "YITP-25-132",
    doi = "10.1007/JHEP10(2025)159",
    journal = "JHEP",
    volume = "10",
    pages = "159",
    year = "2025"
}

@article{Fujiki:2025yyf,
    author = "Fujiki, Kosei and Kanda, Hiroki and Kohara, Michitaka and Takayanagi, Tadashi",
    title = "{Brane cosmology from AdS/BCFT}",
    eprint = "2501.05036",
    archivePrefix = "arXiv",
    primaryClass = "hep-th",
    reportNumber = "YITP-24-180",
    doi = "10.1007/JHEP03(2025)135",
    journal = "JHEP",
    volume = "03",
    pages = "135",
    year = "2025"
}

@article{Shinmyo:2023eci,
    author = "Shinmyo, Kotaro and Takayanagi, Tadashi and Tasuki, Kenya",
    title = "{Pseudo entropy under joining local quenches}",
    eprint = "2310.12542",
    archivePrefix = "arXiv",
    primaryClass = "hep-th",
    reportNumber = "YITP-23-132",
    doi = "10.1007/JHEP02(2024)111",
    journal = "JHEP",
    volume = "02",
    pages = "111",
    year = "2024"
}

@article{Kanda:2023jyi,
    author = "Kanda, Hiroki and Kawamoto, Taishi and Suzuki, Yu-ki and Takayanagi, Tadashi and Tasuki, Kenya and Wei, Zixia",
    title = "{Entanglement phase transition in holographic pseudo entropy}",
    eprint = "2311.13201",
    archivePrefix = "arXiv",
    primaryClass = "hep-th",
    reportNumber = "YITP-23-148",
    doi = "10.1007/JHEP03(2024)060",
    journal = "JHEP",
    volume = "03",
    pages = "060",
    year = "2024"
}

@article{Gu:2017njx,
    author = "Gu, Yingfei and Lucas, Andrew and Qi, Xiao-Liang",
    title = "{Spread of entanglement in a Sachdev-Ye-Kitaev chain}",
    eprint = "1708.00871",
    archivePrefix = "arXiv",
    primaryClass = "hep-th",
    doi = "10.1007/JHEP09(2017)120",
    journal = "JHEP",
    volume = "09",
    pages = "120",
    year = "2017"
}

@article{Chen:2020wiq,
    author = "Chen, Yiming and Qi, Xiao-Liang and Zhang, Pengfei",
    title = "{Replica wormhole and information retrieval in the SYK model coupled to Majorana chains}",
    eprint = "2003.13147",
    archivePrefix = "arXiv",
    primaryClass = "hep-th",
    doi = "10.1007/JHEP06(2020)121",
    journal = "JHEP",
    volume = "06",
    pages = "121",
    year = "2020"
}

@article{Wang:2023vkq,
    author = "Wang, Hanteng and Liu, Chang and Zhang, Pengfei and Garc\'\i{}a-Garc\'\i{}a, Antonio M.",
    title = "{Entanglement Transition and Replica Wormhole in the Dissipative Sachdev-Ye-Kitaev Model}",
    eprint = "2306.12571",
    archivePrefix = "arXiv",
    primaryClass = "quant-ph",
    month = "6",
    year = "2023"
}

@article{Czech:2023rbh,
    author = "Czech, Bartlomiej and Shuai, Sirui and Tang, Haifeng",
    title = "{Entropies and reflected entropies in the Hayden-Preskill protocol}",
    eprint = "2310.16988",
    archivePrefix = "arXiv",
    primaryClass = "hep-th",
    doi = "10.1007/JHEP02(2024)040",
    journal = "JHEP",
    volume = "02",
    pages = "040",
    year = "2024"
}

@article{Li:2021dmf,
    author = "Li, Tianyi and Yuan, Ma-Ke and Zhou, Yang",
    title = "{Defect extremal surface for reflected entropy}",
    eprint = "2108.08544",
    archivePrefix = "arXiv",
    primaryClass = "hep-th",
    doi = "10.1007/JHEP01(2022)018",
    journal = "JHEP",
    volume = "01",
    pages = "018",
    year = "2022"
}

@article{Liu:2026ruv,
    author = "Liu, Yuxuan and Ling, Yi and Xian, Zhuo-Yu",
    title = "{Boundary mutual information in double holography}",
    eprint = "2602.12627",
    archivePrefix = "arXiv",
    primaryClass = "hep-th",
    month = "2",
    year = "2026"
}

@article{Hartnoll:2008vx,
    author = "Hartnoll, Sean A. and Herzog, Christopher P. and Horowitz, Gary T.",
    title = "{Building a Holographic Superconductor}",
    eprint = "0803.3295",
    archivePrefix = "arXiv",
    primaryClass = "hep-th",
    reportNumber = "NSF-KITP-08-38, PUPT-2261",
    doi = "10.1103/PhysRevLett.101.031601",
    journal = "Phys. Rev. Lett.",
    volume = "101",
    pages = "031601",
    year = "2008"
}

@article{Albash:2012pd,
    author = "Albash, Tameem and Johnson, Clifford V.",
    title = "{Holographic Studies of Entanglement Entropy in Superconductors}",
    eprint = "1202.2605",
    archivePrefix = "arXiv",
    primaryClass = "hep-th",
    doi = "10.1007/JHEP05(2012)079",
    journal = "JHEP",
    volume = "05",
    pages = "079",
    year = "2012"
}

@article{Fan:2013wba,
    author = "Fan, ZhongYing",
    title = "{Holographic superconductors with hidden Fermi surfaces}",
    eprint = "1311.4110",
    archivePrefix = "arXiv",
    primaryClass = "hep-th",
    month = "11",
    year = "2013"
}

@article{Kuang:2014kha,
    author = "Kuang, Xiao-Mei and Papantonopoulos, Eleftherios and Wang, Bin",
    title = "{Entanglement Entropy as a Probe of the Proximity Effect in Holographic Superconductors}",
    eprint = "1401.5720",
    archivePrefix = "arXiv",
    primaryClass = "hep-th",
    doi = "10.1007/JHEP05(2014)130",
    journal = "JHEP",
    volume = "05",
    pages = "130",
    year = "2014"
}

@article{Garcia-Garcia:2015emb,
    author = "Garc{\'\i}a-Garc{\'\i}a, Antonio M. and Romero-Berm{\'u}dez, Aurelio",
    title = "{Conductivity and entanglement entropy of high dimensional holographic superconductors}",
    eprint = "1502.03616",
    archivePrefix = "arXiv",
    primaryClass = "hep-th",
    doi = "10.1007/JHEP09(2015)033",
    journal = "JHEP",
    volume = "09",
    pages = "033",
    year = "2015"
}

@article{Wang:2023nes,
    author = "Wang, Dong and Qiao, Xiongying and Wang, Mengjie and Pan, Qiyuan and Lai, Chuyu and Jing, Jiliang",
    title = "{Holographic entanglement entropy and subregion complexity for excited states of holographic superconductors}",
    eprint = "2301.00513",
    archivePrefix = "arXiv",
    primaryClass = "hep-th",
    doi = "10.1016/j.nuclphysb.2023.116223",
    journal = "Nucl. Phys. B",
    volume = "991",
    pages = "116223",
    year = "2023"
}

\appendix
\section{Calculations}\label{app_cal}
\subsection{boundary time}\label{app_bdyt}
The boundary time is
\begin{align}
  t&=
\int_0^{z_t}dz\,\frac{E z}{f(z)\sqrt{f(z)-\frac{z^2}{z_t^2}f(z_t)}}  \nonumber \\
 &=z_h \tanh ^{-1}\left(\frac{z_t}{\sqrt{z_t^2-z_h^2}}\right) =: z_h \tanh ^{-1}x \nonumber\\
 &= \frac{z_h}{2} \log \frac{x+1}{x-1}.
\end{align}
Thus, one obtains
\begin{equation}
    e^{2t/z_h}=\frac{x+1}{x-1}.
\end{equation}
With this expression, one further has
\begin{equation}
    \cosh{\left(\frac{t}{z_h}\right)}=\frac{e^{t/z_h}+e^{-t/z_h}}{2}=\frac{1}{2}\left(\sqrt{\frac{x+1}{x-1}}+\sqrt{\frac{x-1}{x+1}}\right)=\frac{x}{\sqrt{x^2-1}}=\frac{z_t}{z_h}.
\end{equation}
\subsection{variation of the island}\label{app_vi}

Variation of the island on the brane profile gives
\begin{align}
\delta x_p&=\tilde{x}(z_p+\delta z_p)  -x(z_p) \\
          &=\left[x(z_p+\delta z_p) +\delta x(z_p+\delta z_p) \right]-x(z_p)\\
          &=x(z_p)+x'(z_p)\delta z_p + \delta x(z_p+\delta z_p)-x(z_p)\\
          &=x'(z_p)\delta z_p + \delta x(z_p),
\end{align}
with $\tilde{x}$ being a new disconnected surface $\gamma_d$ ended on $z_p+\delta z_p$. Since the island moves along the brane profile as $x_p=X(z_p)$, thus one further has
\begin{equation}
    \delta x_p=X'(z_p)\delta z_p
\end{equation}
Finally, the variation of the island gives
\begin{equation}
    \delta x(z_p)=\left[X'(z_p) -x'(z_p)\right]\delta z_p.
\end{equation}

\subsection{The expressions of $z_*(\theta)$ and $z_p(\theta)$ for analytical solutions}\label{app_para}
For analytical solutions, the transversality condition \eqref{eq:bcs2} reduces to
\begin{equation}\label{eq:th-zpzs}
    \tan^2\theta=\frac{z_p^2 \left(z_*^2-z_h^2\right)}{z_h^2 (z_p^2-z_*^2)}.
\end{equation}
Substitute it into the LHS of the intersection condition \eqref{eq:bcsI}, one has
\begin{equation}
    X(z_p)=\frac{z_h}{2}\log \frac{1-\mathcal{C}}{1+\mathcal{C}}=-z_h \tanh^{-1}\mathcal{C},
\end{equation}
where $$\mathcal{C}=\sqrt{\frac{z_p^2-z_*^2}{z_p^2-z_h^2}}.$$

Then, the entire intersection condition can be expressed as
\begin{align}
    -2 \tanh^{-1}\mathcal{C}&=\frac{x_0}{z_h}  -\tanh^{-1}\frac{z_*}{z_h} \\
    \mathcal{C}&= \tanh \mathcal{P},\label{eq:CP}
\end{align}
where $$\mathcal{P}=\frac{-1}{2}\frac{x_0}{z_h}  +\frac{1}{2}\tanh^{-1}\frac{z_*}{z_h}.$$
From \eqref{eq:CP}, we obtain the expression of $z_p(z_*)$ as
\begin{equation}\label{eq:zp-zs}
    z_p^2=\frac{z_*^2-z_h^2\tanh^2\mathcal{P}}{1-\tanh^2\mathcal{P}}.
\end{equation}
Substitute \eqref{eq:zp-zs} back into \eqref{eq:th-zpzs}, one finally finds
\begin{equation}\label{eq:th-zs}
    \cos\theta=\frac{z_h}{z_*}\tanh \mathcal{P}.
\end{equation}

In practice, the parameter we actually deal with is the electric charge $Q$, or equivalently the angle $\theta$, therefore, it is more convenient to solve the relations $z_*=z_*(\theta)$ and $z_p=z_p(\theta)$ from \eqref{eq:th-zs} and \eqref{eq:zp-zs}.

To obtain an explicit relation for $z_*=z_*(\theta)$, we must rationalize \eqref{eq:th-zs}. First, we introduce two auxiliary variables as
\begin{equation}
\alpha \equiv \tanh\left(\frac{1}{2}\tanh^{-1}\frac{z_*}{z_h}\right),
\qquad
\beta \equiv \tanh\frac{x_0}{2z_h}.
\end{equation}
By using the double-angle identity, one finds
\begin{equation}
    \frac{z_*}{z_h}=\tanh\left(2\, \frac{1}{2}\tanh^{-1}\frac{z_*}{z_h}\right) = \frac{2 \alpha}{1+\alpha^2},
\end{equation}
Similarly, by applying the difference identity, one further finds
\begin{equation}
    \tanh \mathcal{P}=\tanh\left(\tanh^{-1}\alpha-\tanh^{-1}\beta\right) = \frac{\alpha - \beta}{1 - \alpha \beta}.
\end{equation}

Substituting these two expressions back into \eqref{eq:th-zs} yields
\begin{equation}
\cos\theta = \frac{(1+\alpha^2)(\alpha-\beta)}{2\alpha(1-\alpha \beta)}.
\end{equation}

Multiplying out and collecting terms, we obtain the cubic equation of $\alpha$
\begin{equation}
\alpha^3 - \beta(1 - 2\cos\theta)\alpha^2 + (1 - 2\cos\theta)\alpha - \beta = 0.
\end{equation}

This equation admits a physical solution as
\begin{align}
    \alpha&=\alpha(\theta)\\
    &=\frac{1}{3}\left[\left(\frac{c_1}{2}\right)^{1/3}+c_2 \beta + c_2(c_2 \beta^2-3)\left(\frac{2}{c_1}\right)^{1/3}\right],\nonumber\\
    \text{with} \quad c_2&=1-2 \cos \theta \\
    \text{and}  \quad c_1&=2 \beta ^3 c_2^3-9 \beta  \left(c_2^2-3\right)+3 \sqrt{81 \beta ^2-3 \beta ^2 c_2^4+12 \left(\beta ^4+1\right) c_2^3-54 \beta ^2c_2^2}.
\end{align}
Thus, the turning point is reconstructed as
\begin{equation}\label{eq:zs-th}
z_* =z_*(\theta) = z_h \frac{2 \alpha(\theta)}{1+\alpha(\theta)^2}.
\end{equation}

Next, we will search the relation of $z_p=z_p(\theta)$. Substituting \eqref{eq:th-zs} into \eqref{eq:zp-zs}, one finds
\begin{equation}
    z_p^2 = \frac{z_*^2 \sin^2\theta}{1 - \frac{z_*^2}{z_h^2}\cos^2\theta}.
\end{equation}
Combining \eqref{eq:zs-th}, one finally finds
\begin{equation}
    z_p = z_p(\theta)=\frac{2\alpha(\theta) z_h \sin\theta}{\sqrt{1-2\cos 2\theta\; \alpha(\theta)^2 +\alpha(\theta)^4}}.
\end{equation}

\end{document}